\documentclass[twocolumn,superscriptaddress,prb]{revtex4}
\usepackage{graphicx}

\usepackage{graphicx}                       % standard graphics handler
\usepackage{color}  
\usepackage{subeqnarray}                        % colours for the hyperlinks
\usepackage{amsthm,amsmath,amssymb}

\definecolor{pink}{rgb}{1,0.078,0.57}
\definecolor{green}{rgb}{0,0.7,0.9}

\newcommand{\ket}[1] {\left\vert #1 \right\rangle}
\newcommand{\bra}[1] {\left\langle #1 \right|}
\newcommand{\braket}[2] {\langle #1 | #2 \rangle}

\newcommand{\bQ}{\mathbf{Q}}

\begin{document}

\title{Spin-orbit coupling:  atom versus semiconductor crystal}

\author{Monique Combescot}\email[]{combescot@insp.jussieu.fr}
\affiliation{Sorbonne Universit\'e, CNRS, Institut des NanoSciences de Paris, INSP, 75005-Paris, France}
\author{Shiue-Yuan Shiau}\email[]{shiau.sean@gmail.com}
\affiliation{Physics Division, National Center for Theoretical Sciences,  Hsinchu, 30013, Taiwan}
\author{Valia Voliotis}\email[]{voliotis@insp.jussieu.fr}
\affiliation{Sorbonne Universit\'e, CNRS, Institut des NanoSciences de Paris, INSP, 75005-Paris, France}

\date{\today}
\begin{abstract}
We reconsider a key point in semiconductor physics, the splitting of the valence band states induced by the spin-orbit interaction, through a novel approach which uses neither the group theory formalism, nor the usual $\textbf{L}\cdot\textbf{S}$ formulation valid for atoms but conceptually incorrect for periodic lattices, the angular momenta $\textbf{L}$ and $\textbf{J}$  having no meaning due to the absence of spherical symmetry. We show that for zinc-blende structures,  the valence band  eigenstates resulting from spin-orbit coupling are uniquely determined by: (i) the equivalence of the ($x,y,z$) crystal axes, (ii) the three-fold degeneracy of the valence band. The fact that these two conditions are also fulfilled by  atomic $p$ states allows us to understand why the spin-orbit eigenstates for three-fold atomic and  valence electrons have exactly the same structure, albeit the drastic differences in the potential and electronic symmetries. We also come back to the commonly accepted understanding of the exciton-photon interaction in terms of bright and dark excitons having total angular momenta $J=(1,2)$ respectively and present a simple derivation of this interaction which only relies on spin conservation.

\end{abstract}

\maketitle

\section{Introduction}

Spin-orbit coupling is a relativistic effect that is  easy to study in the atomic physics context \cite{Cohen}.
Regrettably, the same approach is commonly used for semiconductor crystals which do not have the spherical symmetry as do atoms\cite{CardonaYu,Lew,Fishman,Wenckebach}. In practical terms, it means that the orbital angular momentum $\textbf{L}$ associated with the \textit{spherical} harmonics $Y_{\ell,\ell_z}(\theta,\varphi)$, and the total angular momentum $\textbf{J}=\textbf{L}+\textbf{S}$,  have no meaning for electrons in a periodic crystal. Yet, taking the valence band in zinc-blende-like semiconductors as a $\ell=1$ atomic $p$ state leads to the same spin-orbit eigenstates and energy splittings as the ones correctly derived from group theory \cite{Ivchenko}. In view of the physically obscure derivation based on group theory, it is tempting to adopt a pragmatic attitude by forgetting about the inconsistency of using the total angular momentum $(j,j_z)$ as quantum indices to characterize the spin-orbit eigenstates of semiconductor crystals. 
 
\textbf{\textsl{The purpose of this work}} is to understand why the spin-orbit eigenstates for three-fold level in a zinc-blende-like semiconductor crystal with cubic symmetry\cite{Dresselhaus}, have the same structure as the $j=(3/2,1/2)$ atomic spin-orbit eigenstates, without resorting to group theory and the hard-to-grasp classification of the various $\Gamma_n$ bands. Here, we derive the effects of the spin-orbit coupling on crystal structures from scratch, by using only  the fact that the valence band is three-fold and the ($x,y,z$) crystal axes are equivalent in a zinc-blende structure.

 To better understand the deep reason why the atomic procedure incorrectly used for semiconductor crystals still leads to the correct spin-orbit eigenstates, we first rederive the spin-orbit splitting for atoms using two different methods: (i) in the first, standard, method, we from the very first line, make use of  the spherical symmetry of the electrostatic potential $\mathcal{V}(\textbf{r})=\mathcal{V}(r)$ felt by the electrons, as induced by the nucleus. This readily leads us to write the spin-orbit coupling in terms of $\textbf{L}\cdot\textbf{S}$ and then of $\textbf{J}$; (ii) in the second method, we propose a pedestrian approach that does not start with any assumption on the symmetry of the potential. This second method is the one that can be used for semiconductor crystals because their electrostatic potential $\mathcal{V}(\textbf{r})$ is not spherically symmetric, but has the lattice periodicity, $\mathcal{V}(\textbf{r})=\mathcal{V}(\textbf{r}+\textbf{a})$, with $\textbf{a}$ being a lattice vector. We show that the Clebsch-Gordan coefficients which relate \textbf{J} to \textbf{L} states and provide the spin-orbit atomic eigenstates, appear in a natural way for semiconductor valence-band eigenstates having a cubic symmetry. This result is not trivial at all because the electrostatic potentials felt by atomic and semiconductor electrons do not have the same symmetry; moreover, atomic $p$ states have an odd parity while orbital states for valence electrons can be even or odd.
 
Our pedestrian approach provides a transparent way to understand the structure of the semiconductor valence band, and its further coupling to photons, as necessary to correctly predict the very rich pattern of polarization effects that result from band  symmetries. In particular, the spin-orbit coupling does not mix up and down spins in two valence states only, these two states being the ones out of which the two dark excitons are constructed.

The paper is organized as follows: in Sec.~\ref{sec:2}, we recall the microscopic expression of the spin-orbit interaction and  provide a short derivation of the spin-orbit interaction along Thomas' understanding \cite{Thomas}, with details given in Appendix \ref{app:A}. Section \ref{sec:3} deals with atoms. The spin-orbit eigenstates are derived in two different ways, the standard one that uses the total angular momentum $\textbf{J}$ being definitely the smartest. Section \ref{sec:4} deals with semiconductor crystals. We consider three-fold degenerate states with  periodic symmetry and  even or odd parity. We show that the zinc-blende-like semiconductor spin-orbit eigenstates have exactly the same structure as the ones of atomic $p$ states, and that this structure does not depend on the valence state parity. In Sec.~\ref{sec:5}, we provide some key results on spin-orbit eigenstates within the language of group theory. We also provide a physical understanding that \textit{a posteriori} explains the similarity in the spin-orbit eigenstates of atoms and cubic semiconductors. We discuss some implications for two-dimensional materials. We finally discuss the importance of separating the (intrinsic) spin subspace from the subspace in which electrons and photons move. This will in particular avoid the incorrect understanding of the exciton-photon interaction in terms of photons having a spin $S=(\pm1,0)$ and dark excitons having a spin $S=\pm2$. We ultimately conclude.

 \section{Spin-orbit interaction\label{sec:2}}
 The general expression of the spin-orbit interaction for an electron having an electrostatic energy $\mathcal{V}(\textbf{r})$, reads as \cite{Cohen}
  \begin{equation}
\label{1}
H_{so}=\lambda_{so}\left(\overrightarrow{\nabla}\mathcal{V}(\textbf{r})\times \textbf{p}\right)\cdot \textbf{S},
\end{equation}  
  where $\textbf{p}=(\hbar/i)\vec{\nabla}$ is the electron momentum operator and $\textbf{S}=   ( \hbar/2)\vec{\sigma}$ is the electron spin operator, the components of the $\vec{\sigma}$ operators being the Pauli matrices, $(\sigma_x,\sigma_y,\sigma_z)$. The $\lambda_{so}$ prefactor is given by
 \begin{equation}
 \label{2}
 \lambda_{so}=\frac{1}{2 m_{0}^2c^2}\,,
 \end{equation}
 where $m_{0}$ is the free electron mass and $c$ is the speed of light. We wish to stress that the spin-orbit interaction given in Eq.~(\ref{1}) is valid for whatever potential $\mathcal{V}(\textbf{r})$. This general expression reduces to the well-known $\textbf{L}\cdot\textbf{S}$ formula in the case of atoms due to the $\mathcal{V}(\textbf{r})$ spherical symmetry. In the case of crystals having a periodic symmetry, we  must stay with Eq.~(\ref{1}), which is less convenient to handle than the $\textbf{L}\cdot\textbf{S}$ form.

 The physical origin of the $1/2$ factor contained in $\lambda_{so}$  has puzzled the leading physicists of the 1920's for quite a long time\cite{Tomonaga}. Appendix \ref{app:A} presents a detailed derivation of this factor along Thomas' idea\cite{Thomas}. It relies on a succession of different physical effects that we find of interest to  outline below.
  
 \textbf{(1)} An electron with mass $m_0$, charge $e=-|e|$, and spin $\textbf{S}$, has  a magnetic moment $\textbf{M}_S$,
 \begin{equation}
\label{3}  \textbf{M}_S= g_{e}\frac{e}{2m_{0}c}\textbf{S}\,,
\end{equation}
the Land\'e factor $g_{e}$ being equal to 2 for the electron spin.
 
 \textbf{(2)} In an external magnetic field $\textbf{H}_{ext}$, the electron energy associated with its  $\textbf{M}_S$ magnetic moment reads 
 \begin{equation}
\label{4}
-\textbf{M}_{S}\cdot \textbf{H}_{ext}=-g_{e}\frac{e}{2m_{0}c}\textbf{S}\cdot \textbf{H}_{ext}\,.
\end{equation}     

\textbf{(3)} We now consider an electron moving with a velocity $\textbf{v}=\textbf{p}/m_0$ in the laboratory frame F in which exists an electromagnetic field $(\textbf{H}_{ext}, \textbf{E}_{ext})$. The Lorentz transformation gives the magnetic part of this field in a frame F$'$ that moves at a velocity $\textbf{v}$ with respect to the F frame, as
\begin{equation}
\label{5}
\frac{\textbf{H}_{ext}-\frac{\textbf{v}}{c}\times \textbf{E}_{ext}}{\sqrt{1-v^2/c^2}}\simeq \textbf{H}_{ext}+\frac{\textbf{E}_{ext}\times \textbf{v}}{c}\,.
\end{equation}     

\textbf{(4)} In the F$'$ frame, the electron also feels an electric force, which at lowest order in $v/c$ is equal to $e\textbf{E}_{ext}$. So, the charged electron feels an acceleration $\textbf{a}$ given by
\begin{equation}
\label{six}
m_{0}\textbf{a}=e\textbf{E}_{ext}\,.
\end{equation}
Consequently, the frame in which the electron is at rest is not the  F$'$ frame that moves at a constant velocity $\textbf{v}$, but a frame that accelerates. As a result, Eq.~\eqref{5}, which results from a Lorentz transformation valid for constant velocity, does not  give the correct magnetic field felt by the electron.

\textbf{(5)} It is possible to solve this problem by using the Dirac equation\cite{LL}. This equation  gives  relativistic corrections associated with spin, up to the $1/c^2$ order, as
\begin{equation}
\label{7}
- \textbf{S}\cdot \left[\frac{e}{m_{0}c} \textbf{H}_{ext}+\frac{e}{2m_{0}c^2}\textbf{E}_{ext}\times \textbf{v}\right]\,.
\end{equation}       
The second term  corresponds to the spin-orbit interaction. It is absent from the Pauli equation (see Eq.~(\ref{A16})), which only is correct up to the $1/c$ order. 

\textbf{(6)} Actually, it is possible to bypass relativistic quantum theory by using Thomas' argument\cite{Thomas}. It relies on the keen observation that the acceleration of the frame in which the electron is at rest, corresponds to a rotation with respect to the laboratory frame, with an angular precession velocity given by 
\begin{equation}
\label{8}
\mathbf{\Omega}_{acc}=- \frac{\textbf{v}\times \textbf{a}}{2c^2}
\end{equation}
(see Appendix \ref{app:Thomas} for details on this key result).

\textbf{(7)} The derivation then follows by noting that an electron spin $\textbf{S}$ in a magnetic field $\textbf{H}$ along the $z$ direction has an energy $-\textbf{M}_{S}\cdot \textbf{H} \equiv \omega_{H}S_{z}$ that makes it rotate around $\textbf{H}$ with a frequency $\omega_{H}$ which corresponds to an angular precession velocity
\begin{equation}
\label{9}
\mathbf{\Omega}_{H}=- \frac{M_{S}}{S}\textbf{H}\,.
\end{equation}
So, this magnetic field $\textbf{H}$ gives to the spin-$\textbf{S}$ electron a rotational kinetic energy that can be written as $\textbf{S}\cdot \mathbf{\Omega}_{H}$. Conversely, the angular precession velocity $\mathbf{\Omega}_{acc}$ induced by the electron acceleration $\textbf{a}$ in Eq.~\eqref{six} corresponds to an effective magnetic field which gives the spin-$\textbf{S}$ electron an energy
\begin{equation}
\label{10}
\textbf{S}\cdot \mathbf{\Omega}_{acc}=-\frac{1}{2c^2}\textbf{S}\cdot \big(\textbf{v}\times\frac{e\textbf{E}_{ext}}{m_{0}}\big)=\frac{e}{2m_{0}c^2}\textbf{S}\cdot (\textbf{E}_{ext}\times \textbf{v})\,.
\end{equation}

This electron energy has to be added to the energy due to the magnetic field given in Eq.~\eqref{5}. So, the total energy of a spin-$\textbf{S}$ electron moving with a velocity $\textbf{v}$ in a  $(\textbf{H}_{ext}, \textbf{E}_{ext})$ field reads as
\begin{equation}
\label{11}
 -\frac{e}{2m_{0}c} \textbf{S}\cdot\left[g_{e}\textbf{H}_{ext}+g_{e}\frac{\textbf{E}_{ext}\times \textbf{v}}{c} - \frac{\textbf{E}_{ext}\times \textbf{v}}{c}\right]\,.
\end{equation}

Since $g_{e}=2$, while the electrostatic energy $e\textbf{E}_{ext}$ can be written  in terms of the electrostatic potential $\mathcal{V}(\textbf{r})$ as $-e\overrightarrow{\nabla}  \mathcal{V}(\textbf{r})$, the second and third terms of the above equation combine to give the spin-orbit interaction as 
\begin{equation}
-\frac{e}{2m_{0}c^2} \textbf{S}\cdot \textbf{E}_{ext}\times \textbf{v}=\frac{1}{2m_{0}^{2}c^2} \textbf{S}\cdot (\overrightarrow{\nabla} \mathcal{V}(\textbf{r})\times \textbf{p})\,,
\end{equation}
in agreement with Eqs.~(\ref{1}) and (\ref{2}). The $1/2$ prefactor directly follows from $\mathbf{\Omega}_{acc}$ given in Eq.~(\ref{8}). To prove this crucial result is rather lengthy; this is why we have relegated its detailed derivation to Appendix \ref{app:Thomas}.

\section{Spin-orbit splitting for atoms \label{sec:3}}
  In the case of atoms, the  electrostatic potential due to the nuclear charge has spherical symmetry. So, for $\mathcal{V}(\textbf{r})= \mathcal{V}^{(at)}(r)$,  
 \begin{equation}
\label{12}
\overrightarrow{\nabla}\mathcal{V}(\textbf{r})=\frac{\textbf{r}}{r}  \, \frac{d \mathcal{V}^{(at)}(r)}{d r}\,.
\end{equation}
The spin-orbit interaction in Eq.~\eqref{1} then reduces to
 \begin{equation}
\label{13}
H_{so}^{(at)}=\lambda_{so}(r)  \Big(\textbf{r}\times \textbf{p} \Big)\cdot \textbf{S}\,,
\end{equation}
where $ \lambda_{so}(r)$  is a positive scalar that depends on $r$ as
 \begin{equation}
\label{14}
\lambda_{so}(r) = \frac{\lambda_{so}}{r}   \,   \frac{d\mathcal{V}^{(at)}(r)}{dr}     =     \lambda_{so}     \frac{\textbf{r}}{r^2}   \cdot   \overrightarrow{\nabla}\mathcal{V}^{(at)}(r)   \,.             
 \end{equation}
Indeed, the electrostatic potential $\mathcal{V}^{(at)}(r)$ felt by the electron is attractive, that is, negative, being minimum for $r$ small. 

\subsection{Derivation using the standard $\textbf{L}\cdot {\textbf{S}}$ formulation} 
$\bullet$ The standard way to derive the spin-orbit energy splitting of  atomic levels is to note that $(\textbf{r}\times \textbf{p})$  is just $\textbf{L}$ the electron orbital angular momentum\cite{Cohen,Baym,Messiah}. So, Eq.~(\ref{1}) also reads
\begin{equation}
\label{15}
H_{so}^{(at)}=\lambda_{so}(r) \, \textbf{L}\cdot {\textbf{S}}=\frac{1}{2}\lambda_{so}(r) (\textbf{J}^2-\textbf{L}^2-\textbf{S}^2)\,,
\end{equation}
where $\textbf{J}=\textbf{L}+\textbf{S}$ is the total angular momentum of the atomic electron.
This spin-orbit coupling splits the ($2\times 3$)-fold degeneracy of an electron with spin $\pm1/2$ in a  $p$ atomic level, into four-fold and two-fold states which  correspond to $\ket{j,j_z}$ with $j=(3/2,1/2)$ and $-j\leqslant j_{z}\leqslant j$. Since all states are made of $\ell=1$ orbital states and $\textsl{s}=1/2$ spin states, the above equation gives for the two different $j$ values, 
 \begin{equation}
\label{16}
\textbf{L}\cdot\textbf{S}\ket{j,j_z} 
=\frac{\hbar^{2}}{2}\left[j\left(j{+}1\right){-}1(1{+}1){-}\frac{1}{2}\left(\frac{1}{2}{+}1\right)\right]\ket{j,j_z}  \,, 
\end{equation}
the bracket being equal to 1 for $j=3/2$ and to $-2$ for $j=1/2$.
The above equation shows that the $H_{so}^{(at)}$ eigenstates  correspond to the four-fold $\ket{3/2,j_z}$ states and the two-fold $\ket{1/2,j_z}$ states.

$\bullet$ The normalized $\ket{j,j_z}$ states read in terms of the $\ket{\ell_z=(\pm1,0)}\otimes\ket{\textsl{s}_z=\pm1/2}$ states as (see Appendix \ref{derjjz})
\begin{subeqnarray}
\label{23}
 \ket{\frac 3 2,\frac{3\eta}{2}}  &=&  \ket{\eta}  \otimes \ket{\frac{\eta}{2}}  ,  \\
\ket{\frac 3 2,\frac{\eta}{2}}  &=&  \frac{1}{\sqrt{ 3}}\left(\ket{\eta}        \otimes \ket{-\frac{\eta}{2}}+\sqrt{ 2}\ket{0}\otimes\ket{\frac{\eta}{2}}   \right)  , \\
\ket{\frac 1 2,\frac{\eta}{2}}  &=&  \frac{1}{\sqrt{ 3}}\left(\sqrt{ 2}\ket{\eta}        \otimes \ket{-\frac{\eta}{2}}-\ket{0}\otimes\ket{\frac{\eta}{2}} \right).
\slabel{23'}
\end{subeqnarray}
for $\eta=\pm1$. By using Landau-Lifschitz phase factor for the $Y_{\ell,\ell_z}(\theta,\varphi)$ spherical harmonics\cite{LLMQ}, the $\ket{\ell_z=(\pm1,0)}$ states read in terms of the $\ket{\lambda=(x,y,z)}$ states as
\begin{subeqnarray}
\label{18}
\ket{\pm1} & = & \frac{\mp i\ket{x}+\ket{y}}{\sqrt{2}}  \,,
 \\
\ket{0} & = & i\ket{z}\,.
\end{subeqnarray}

The orbital part of the atomic state ($n,\ell,\ell_z)$, where $n$ is the principal quantum number, corresponds to $R_{n,\ell}(r)\,\,Y_{\ell,\ell_z}(\theta,\varphi)$, which gives, for the atomic $p$ states labeled by $\lambda$,
\begin{equation}
\label{17}
\braket{\textbf{r}}{\lambda} =i\sqrt{\frac{3}{4\pi}}\frac{\lambda}{r} \,\,R_{n,\ell=1}(r)\,.
\end{equation} 
Note that the state norm
\begin{equation}
\label{19}
\braket{\lambda}{\lambda}=\frac{3}{4\pi}\int \mathrm{d}^{3}r \,\frac{\lambda^{2}}{r^2}  |R_{n,1}(r)|^{2}=\frac{1}{4\pi}\int \mathrm{d}^{3}r  |R_{n,1}(r)|^{2}
\end{equation} 
does not depend on $\lambda$; so, this norm is also equal to the norms $\braket{\ell_z}{\ell_z}$ with $\ell_{z}=(0,\pm1)$.

Using Eq.~(\ref{16}), this gives the eigenvalues of the spin-orbit operator as
\begin{eqnarray}
\label{21}
\lefteqn{\frac{\bra {j,j_z} H_{so}^{(at)}\ket{j,j_z}}{\braket{j,j_z}{j,j_z}} } \\ 
&&=\mathcal{O}_j \,\, \frac{\hbar^{2}} {2     } \,\,  \frac{\int \mathrm{d}^{3}r  \,\lambda_{so}(r)|R_{n,1}(r)|^{2}}   {\int \mathrm{d}^{3}r  \, |R_{n,1}(r)|^{2}} 
\equiv  \mathcal{O}_j \,\Lambda_{so}^{(at)} \,,
 \nonumber
 \end{eqnarray}
 with $\mathcal{O}_j $ equal to $1$ for $j=3/$2 and to $-2$ for $j=1/2$.

All this shows that the  spin-orbit interaction, $H_{so}^{(at)}$, induces a splitting between the $(2\times3)$-fold atomic $p$ states into $(4+2)$ eigenstates having an energy difference $3 \Lambda_{so}^{(at)}$.
 Since $\Lambda_{so}^{(at)}$ is positive, the four $j=3/2$ states have an energy increase equal to $\Lambda_{so}^{(at)}$, while the other two $j=1/2$ states have an energy decrease equal to $-2\Lambda_{so}^{(at)}$, making the average energy unchanged by the spin-orbit interaction.

\subsection{Pedestrian approach \label{sec:3b}}
$\bullet$ The fact that $\mathcal{V}(\textbf{r})$  depends only on $r$  readily leads to $\textbf{r}\times \textbf{p}$, 
that is, $\textbf{L}$, and ultimately $\textbf{J}$. However, for electrons in a periodic lattice, $\mathcal{V}(\textbf{r})$ is periodic, $\mathcal{V}(\textbf{r})=\mathcal{V}(\textbf{r}+\textbf{a})$; so, we must find a procedure to derive spin-orbit eigenstates that does not use the orbital angular momentum $\textbf{L}$ as in Eq.~\eqref{13}. To this end, we introduce the two vectors 
\begin{equation}
\label{26}
\textbf{W}(\textbf{r})=\overrightarrow{\nabla} \mathcal{V}(\textbf{r})\,, \qquad 
\vec{\mathcal{L}}
(\textbf{r})=\textbf{W}(\textbf{r})\times \textbf{p}\,.
\end{equation} 
The spin-orbit interaction given in Eq.~\eqref{1} then reads
\begin{equation}
\label{27}
H_{so}=\lambda_{so}   \,\,   \vec{\mathcal{L}} \cdot  \textbf{S}= \lambda_{so} \Big (\frac{\mathcal{L}_{+1} S_{-1} +    \mathcal{L}_{-1}  S_{+1}}{2}  + \mathcal{L}_z  S_z \Big)\,,
\end{equation}
with $\mathcal{L}_{\pm1}=\mathcal{L}_x    \pm i  \mathcal{L}_y $, and similarly for $S_{\pm 1}$.

 As $S_{\pm 1} \ket{\eta/2}= (\hbar/2) \big(1\mp\eta \big)   \ket{-\eta/2} $ while $S_z\ket{\eta/2}= (\eta\hbar/2) \ket{\eta/2}$,  the spin part of $H_{so}$ leads to 
\begin{equation}
\label{28}
H_{so} \ket{\eta/2}=   \lambda_{so}  \frac{\hbar}{2}  \Big (  \eta     \mathcal{L}_z   \ket{\eta/2}  +  \mathcal{L}_ \eta   \ket{-\eta/2}  \Big) \,.            
\end{equation}
This evidences a key point of the derivation: the orbital operator associated with a spin flip is $\mathcal{L}_{\eta}$.

$\bullet$ Next, we turn to the orbital part. The wave functions of the $\vec{\mathcal{L}}\ket{\lambda}$ states for $\braket{\textbf{r}}{\lambda}$ defined in Eq.~\eqref{17}, appear as
\begin{eqnarray}
\label{29}
\bra{\textbf{r}} \mathcal{L}_{x} \ket{x}
& = & \frac{\hbar}{i}\left( W_{y} \frac{\partial}{\partial z} - W_{z} \frac{\partial}{\partial y}\right)\braket{\textbf{r}}{x}     \\
 & = &  \frac{\hbar}{i}\frac{xyz}{r}f'\left[\frac{W_{y}}{y}-\frac{W_{z}}{z}\right] ,\nonumber 
\end{eqnarray}
where  $f(r)=i\sqrt{3/4\pi}\,\,R_{n,1}(r)/r$ and $f'=df(r)/dr$.
 In the same way,
\begin{subeqnarray}
\label{30}
\bra{\textbf{r}} \mathcal{L}_{y} \ket{x} \!\!& = & \frac{\hbar}{i} \left(fW_{z}+\frac{x^{2}z}{r}f' \left[\frac{W_{z}}{z}-\frac{W_{x}}{x}\right]\right),\\
 \bra{\textbf{r}} \mathcal{L}_{z} \ket{x} \!\!& = & \!\! \frac{\hbar}{i} \left(-fW_{y}+\frac{x^{2}y}{r}f' \left[\frac{W_{x}}{x}-\frac{W_{y}}{y}\right]\right).
\end{subeqnarray}

$\bullet$ For a spherical potential, $\mathcal{V}(r)=\mathcal{V}^{(at)}(r)$, the components of  $\textbf{W}(\textbf{r})$ are such that 
\begin{equation}
\label{31}
\frac{W_{x}^{(at)}}{x}=\frac{W_{y}^{(at)}}{y}=\frac{W_{z}^{(at)}}{z}=\frac{1}{r}\frac{\mathrm{d}\mathcal{V}^{(at)}(r)}{\mathrm{d}r}=\frac{\lambda_{so}(r)}{\lambda_{so}}\,.
\end{equation}
This makes all $f'$ factors in $\bra{\textbf{r}} \mathcal{L}_{\lambda'}^{(at)} \ket{\lambda}$ equal to zero. So, we are left with $\bra{\textbf{r}} \mathcal{L}_{x}^{(at)} \ket{x}=0$, and $\bra{\textbf{r}} \mathcal{L}_{y}^{(at)} \ket{x}$ and $\bra{\textbf{r}} \mathcal{L}_{z}^{(at)} \ket{x}$ respectively proportional to $f W_z^{(at)}$ and $f W_y^{(at)}$;
the other terms are obtained from cyclic permutations.\

Next, we note that $f W_z^{(at)}= \braket{\textbf{r}}{z}\lambda_{so}(r)/\lambda_{so}$, but 
 we cannot readily conclude that $\mathcal{L}_{y}^{(at)}$ acting on $\ket{x}$ gives $\ket{z}$ because $\lambda_{so}(r)$ depends on $r$. Actually, what we need to do is  diagonalize the spin-orbit interaction  $H_ {so}^{(at)}$  in the degenerate subspace $ \ket{\lambda}  \otimes \ket{\eta/2}$. In this subspace, 
 
 \begin{eqnarray}
\label{33}
 \Big(\vec { \mathcal{L}}^{(at)}\ket{\lambda}\Big)_{proj}= \sum _{\lambda'=(x,y,z)} \frac { \ket{\lambda'}\bra{\lambda'}}  { \braket{\lambda'}{\lambda'}}\vec { \mathcal{L}}^{(at)}\ket{\lambda}\,.                                
\end{eqnarray}

Using Eq.~\eqref{31}, we find that $ \bra{\lambda'} \mathcal{L}_{\lambda''}^{(at)} \ket{\lambda}$ is equal to zero when any two of $(\lambda,\lambda',\lambda'')$ are the same, like 
$\bra{\lambda} \mathcal{L}_{\lambda''}^{(at)} \ket{\lambda}=0$. The non-zero terms follow by cyclic permutations from 
\begin{equation}
\label{35}
\bra{z} \mathcal{L}_{y}^{(at)} \ket{x}%=\frac{3\hbar}{4\pi i}   \int \mathrm{d}^{3}r  \,\,  |R_{n,1}(r)|^{2}    \frac{z^2}{r^2}   \frac{W_z^{(at)}}{z} 
\equiv \frac{\hbar}{i}           \mathcal{W}^{(at)}  \,,                        
\end{equation}
where $\mathcal{W}^{(at)}$  reads, using Eqs.~(\ref{21}) and (\ref{31}),
\begin{equation}
\label{36}
\mathcal{W}^{(at)}= \frac{1}{4\pi}\int \mathrm{d}^{3}r  \,  |R_{n,1}(r)|^{2}     \, \frac{\lambda_{so}(r)}{\lambda_{so}}=\frac{2}{\hbar^2}\frac{\Lambda_{so}^{(at)}}{\lambda_{so}}\braket{\lambda}{\lambda}\,.
\end{equation}
This leads, for the ($\ket{\pm1},\ket{0}$) states defined in Eq.~(\ref{18}), to
\begin{eqnarray}
 \bra{0} \mathcal{L}_{z}^{(at)} \ket{0}\!& = &\! \bra{\eta'} \mathcal{L}_{z}^{(at)} \ket{0}=\bra{\eta'} \mathcal{L}_{\pm1}^{(at)} \ket{\eta}=0,\nonumber
   \\
   \bra{\eta'} \mathcal{L}_{z}^{(at)} \ket{\eta}\!& = &\!\hbar  \, \mathcal{W}^{(at)}\frac{ \eta'+\eta}  {2},\label{37}
      \\
    \bra{\eta'} \mathcal{L}_{\eta}^{(at)} \ket{0}\!& = &\!\hbar  \, \mathcal{W}^{(at)}\frac{1+ \eta'\eta}  {\sqrt{2}},\nonumber
\end{eqnarray}
which yield, with the help of Eq.~\eqref{33},
\begin{eqnarray}
  & 0\!\!\! & {=}  \Big(\mathcal{L}_z^{(at)}\ket{0}\Big)_{proj} ,\nonumber \\      
   &  \displaystyle\frac{2}{\hbar} \frac{\Lambda_{so}^{(at)}}{\lambda_{so}} \frac{1-\eta' \eta}{\sqrt{2}}\ket{0}\!\!\! & {=}\Big( \mathcal{L}_{\eta'}^{(at)}\ket{\eta}\Big)_{proj}  , \label{38}  \\
 & \displaystyle \frac{2}{\hbar} \sqrt{2}  \,
         \frac{\Lambda_{so}^{(at)}}{\lambda_{so}} \ket{\eta}\!\!\! & {=}\Big(\mathcal{L}_{\eta}^{(at)}\ket{0}\Big)_{proj}{=} \sqrt{2}\eta \Big(\mathcal{L}_{z}^{(at)}\ket{\eta}\Big)_{proj}. \nonumber
\end{eqnarray}
Using Eq.~(\ref{28}), we end with $H_{so}^{(at)}$ acting in the degenerate $ \ket{\lambda}  \otimes \ket{\eta/2}$ subspace as\begin{eqnarray}
\label{39}
H_{so}^{(at)} \ket{\eta'}  \otimes \ket{\frac \eta 2}&=&\Lambda_{so}^{(at)}\Big( \eta\eta' \ket{\eta'}  \otimes \ket{\frac \eta 2} 
      \\
 & & + \frac{1-\eta\eta'  } {\sqrt{2}}  \ket{0}  \otimes \ket{-\frac \eta 2}\Big)\,,
   \nonumber   \\
H_{so}^{(at)} \ket{0}  \otimes \ket{\frac \eta 2}&=&\sqrt{2}\Lambda_{so}^{(at)} \ket{\eta}  \otimes \ket{-\frac \eta 2}\,.
\end{eqnarray}

Equation (\ref{39}) taken for $\eta=\eta'$ readily shows that the $ \ket{\eta}  \otimes \ket{\eta/2}$ states for $\eta=\pm1$ are eigenstates of $H_{so}^{(at)}$ with eigenvalue $\Lambda_{so}^{(at)}$, in agreement with Eq.~(\ref{21}). This equation also shows that the other four eigenstates are linear combinations of $\ket{\eta}  \otimes \ket{-\eta/2}$ and $\ket{0}  \otimes \ket{\eta/2}$, namely

\begin{equation}
\label{40}
\ket{\Psi}=A\ket{0}  \otimes \ket{\frac \eta 2}   + B\ket{\eta}  \otimes \ket{-\frac \eta 2}\,.
\end{equation} 
By noting that
\begin{eqnarray}
\label{41}
H_{so}^{(at)}\ket{\Psi}&=&\Lambda_{so}^{(at)} \Big\{A\sqrt{2}\ket{\eta}  \otimes \ket{-\frac \eta 2}\\
&&+B\big(-\ket{\eta}  \otimes \ket{-\frac \eta 2}
 +\sqrt{2}\ket{0}  \otimes \ket{\frac \eta 2}\big)\Big\}\,, \nonumber
  \end{eqnarray}
  we see that $\ket{\Psi}$ is eigenstate of $H_{so}^{(at)}$ with the eigenvalue $\gamma\Lambda_{so}^{(at)}$, provided that $\gamma B=-B+A\sqrt{2}$ and $\gamma A=B\sqrt{2}$. Non-zero $(A,B)$ values, solution of these two linear homogeneous equations, impose  $\gamma^2+\gamma-2=0$, which gives $\gamma=(1,-2)$ and $A=(B\sqrt{2},-B/\sqrt{2})$. It is easy to check that the corresponding four $\ket{\Psi}$ states are $\ket{j=1/2,\eta   /2} $ and $\ket{j=3/2,\eta/2} $ given in Eq.~(\ref{23}), with the spin-orbit shifts given in Eq.~(\ref{21}). \
  
 Obviously, this approach is not as smart as the one that uses angular momentum operators $\textbf{J}$ and $\textbf{L}$. However, since $\textbf{J}$ has no meaning for electrons in a periodic lattice, an  approach that does not rely on $\textbf{J}$ and $\textbf{L}$ is mandatory. Let us now see what this second approach gives for the spin-orbit eigenstates of semiconductor crystals with cubic symmetry.
  
\section{Spin-orbit splitting for semiconductor crystals\label{sec:4}}

The first problem is to understand the consequences of having the brackets in Eqs.~(\ref{29}) and (\ref{30}) different from zero, as for potentials without spherical symmetry. The second problem is to understand the consequences of the valence state parity because atomic $p$ states are odd, while for zinc-blende structures, the three-fold valence orbital states do not have a defined parity since these structures do not possess inversion symmetry\cite{LK,BirPikus,CardonaYu,Ivchenko}.

\subsection{Periodic potential}
 To derive the effect of the spin-orbit interaction given in Eq.~\eqref{1} for a periodic crystal, we fundamentally follow the procedure used in Sec.\ref{sec:3b}, that is, we write the spin-orbit interaction $H_{so}=\lambda_{so} \vec{\mathcal{L}}(\textbf{r})\cdot \textbf{S}$ as in Eq.~\eqref{27} with $\vec{\mathcal{L}}(\textbf{r})$ defined in Eq.~\eqref{26}, and we handle the periodicity of the potential through its expansion
\begin{equation}
\label{43}
\mathcal{V}(\textbf{r})=\displaystyle \sum_{\textbf{Q}}\mathcal{V}_{\textbf{Q}}e^{i\textbf{Q}\cdot\textbf{r}}
\end{equation}
on reciprocal lattice vectors $\textbf{Q}$ that are such that $e^{i\textbf{Q}\cdot\textbf{a}} =1$,\textit{ i.e.}, $\textbf{Q}$ vectors quantized in $2\pi /|\textbf{a}|$, in order  to fulfill $\mathcal{V}(\textbf{r})=\mathcal{V}(\textbf{r}+\textbf{a})$. This gives
\begin{equation}
\label{44}
\textbf{W}(\textbf{r})=\overrightarrow{\nabla} \mathcal{V}(\textbf{r}) = i \displaystyle \sum_{\textbf{Q}} \textbf{Q}\, \mathcal{V}_{\textbf{Q}}e^{i\textbf{Q}\cdot\textbf{r}}\,.
\end{equation}

\subsection{Valence and conduction states }

$\bullet$ The $s$ atomic levels are non-degenerate, with an even wave function $\psi_{n00}(r)$, while the $p$ atomic  levels are three-fold, with an odd  wave function reading as $\psi_{n,1,\lambda}(r)$ for $\lambda=(x, y, z)$, or any linear combination, $(x,y,z)$ being arbitrary orthogonal axes due to the spherical symmetry of the problem. \

In a crystal, there are two types of non-degenerate states improperly called $s$: one type is even as $s$ atomic levels, the other type is odd. Similarly, there are two types of degenerate states improperly called $p$; they both are three-fold but one type is odd like $p$ atomic  levels, while the other type is even. Due to the lack of defined parity, the semiconductor conduction and valence bands are linear combinations of these even and odd states.

$\bullet$ Electrons in a periodic crystal are characterized by a momentum $\textbf{k}$ and a band index $n$. Their wave functions read 
\begin{equation}
\label{45}
\braket{\textbf{r}}{n;\textbf{k}}=\frac{e^{i\textbf{k} \cdot\textbf{r}}}{L^{3/2}}u_{n;\textbf{k}}(\textbf{r})
\end{equation}
for a sample volume $L^3$. This wave function contains a $e^{i\textbf{k} \cdot\textbf{r}}/L^{3/2}$ part  with momentum \textbf{k} quantized in $2\pi /L$, that just corresponds to a plane wave in free space, and a $u_{n;\textbf{k}}(\textbf{r})$ part that has the lattice periodicity,  $u_{n;\textbf{k}}(\textbf{r})=u_{n;\textbf{k}}(\textbf{r}+\textbf{a})$.

We look for the shift of the band extrema induced by the spin-orbit interaction. In a GaAs-like direct gap semiconductor\cite{Kireev,Klingshirn}, these extrema are located at $\textbf{k}=\textbf{0}$, called  $\Gamma$ point.  
To handle the $u_{n;\textbf{k}=\textbf{0}}(\textbf{r})$ periodicity, we do as for $\mathcal{V}(\textbf{r})$, that is, we expand it on the reciprocal lattice vectors, $e^{i\textbf{K} \cdot \textbf{a}}=1$. Equation \eqref{45} then gives
\begin{equation}
\label{46}
\braket{\textbf{r}}{n;\textbf{k}=\textbf{0}}=\frac{1}{L^{3/2}} \sum_\textbf{K}  \mathcal{U}_{n;\textbf{K}}
e^{i\textbf{K}\cdot\textbf{r}}\,.
\end{equation}

$\bullet$ Next, we note that the valence states $(n=v)$ are characterized by an additional three-fold index $\lambda$  that can still be labeled as $(x,y,z)$, but $(x,y,z)$ are now  the three axes of the cubic crystal at hand.

\textbf{(i)} Odd valence states are such that $\braket{\textbf{r}}{\lambda, v;\textbf{0}}=-\braket{-\textbf{r}}{\lambda, v;\textbf{0}}$. In the reciprocal space, this implies $\mathcal{U}_{\lambda, v;\textbf{K}}=-\mathcal{U}_{\lambda, v;-\textbf{K}}$ that can be written as  
\begin{equation}
\label{47}
\mathcal{U}_{\lambda, v;\textbf{K}}=K_\lambda  G_o(K)\,,
\end{equation}

where $K=|\textbf{K}|$.

\textbf{(ii)} Even valence states are such that $\braket{\textbf{r}}{\lambda, v;\textbf{0}}=\braket{-\textbf{r}}{\lambda, v;\textbf{0}}$. This implies $\mathcal{U}_{\lambda, v;\textbf{K}}=\mathcal{U}_{\lambda, v;-\textbf{K}}$ that can be written as
\begin{equation}
\label{48}
\mathcal{U}_{\lambda, v;\textbf{K}}= \frac{K_xK_yK_z}{K_\lambda}G_e(K)\,.
\end{equation}

\subsection{$\vec{\mathcal{L}}(\textbf{r})$ in the degenerate subspace }
To get the spin-orbit interaction acting on the three-fold degenerate states $\ket{\lambda, v;\textbf{0}}$, we first have to calculate the matrix elements of  $\mathcal{\vec{L}}(\textbf{r})$, defined in Eq.~\eqref{26}, in this subspace.

By noting that 
\begin{equation}
\label{49}
\bra{\textbf{r}}\textbf{p} \ket {\lambda, v;\textbf{0}}= \frac{\hbar}{i} \overrightarrow{\nabla} 
\braket{\textbf{r}} {\lambda, v;\textbf{0}} = \frac{\hbar}{L^{3/2}} \sum_\textbf{K} \textbf{K} \,   \mathcal{U}_{\lambda, v;\textbf{K}}\,e^{i\textbf{K}\cdot\textbf{r}}\,,
\end{equation}
we readily get, for $\textbf{W}(\textbf{r})$ given in Eq.~(\ref{44}),
\begin{eqnarray}
\label{50}
\bra{\textbf{r}} \vec{\mathcal{L}}(\textbf{r}) \ket {\lambda, v;\textbf{0}}&=&\bra{\textbf{r}}\textbf{W}(\textbf{r})\times \textbf{p}\ket {\lambda, v;\textbf{0}}
 \\
 &=& \frac{i \hbar}{L^{3/2}}\sum_{\textbf{Q},\textbf{K}}\big (\textbf{Q}\times \textbf{K}\big ) \mathcal{V}_\textbf{Q} \, \mathcal{U}_{\lambda, v;\textbf{K}}\,e^{i(\textbf{K}+\textbf{Q})\cdot\textbf{r}} \,. 
  \nonumber 
\end{eqnarray}

Next, as in Eq.~(\ref{33}), we look for $\vec{\mathcal{L}}(\textbf{r})$ acting in the degenerate subspace $\ket{\lambda,v;\textbf{0}}$, namely
\begin{equation}\label{51'}
\vec {\mathcal{L}}(\textbf{r})\ket{\lambda, v;\textbf{0}}_{proj}=\!\!\sum_{\lambda'=(x,y,z)} \frac {\ket{\lambda',v;\textbf{0}}\bra{\lambda',v;\textbf{0}}}  {\braket{\lambda',v;\textbf{0}}{\lambda',v;\textbf{0}}}\vec {\mathcal{L}}(\textbf{r})\ket{\lambda, v;\textbf{0}}  \,.                             
\end{equation}
The above matrix element follows from Eq.~(\ref{50}) as 
\begin{eqnarray}
 \lefteqn{\bra{\lambda',v;\textbf{0}} \vec {\mathcal{L}}(\textbf{r})\ket{\lambda, v;\textbf{0}}  = i\hbar   \sum _\textbf{Q} \! \sum _{\textbf{K}'\textbf{K}}\!  \big(\textbf{Q}\times \textbf{K}\big) \mathcal{V}_\textbf{Q}
  }\hspace{1cm}
   \nonumber \\
 &  &\times\mathcal{U}^*_{\lambda',v;\textbf{K}'} \mathcal{U}_{\lambda, v;\textbf{K}}\int   \frac {\mathrm{d}^{3}r}  {L^3} e^{i(\textbf{K}+\textbf{Q}-\textbf{K}')\cdot\textbf{r}}\label{52}  \\
 & =& i\hbar   \sum_{\textbf{K}',\textbf{K}}  \mathcal{V}_{\textbf{K}'-\textbf{K}}
 \mathcal{U}^*_{\lambda',v;\textbf{K}'} \mathcal{U}_{\lambda, v;\textbf{K}}
 \,\,  ( \textbf{K}'{-}\textbf{K})\times \textbf{K}\,, \nonumber
 \end{eqnarray}
with $(\textbf{K}'-\textbf{K})\times \textbf{K}$ reducing to $  \textbf{K}'{\times} \textbf{K}$.
 
To go further, we note that the electrostatic potential of an electron in a cubic crystal with $(x,y,z)$ axes, fulfills
\begin{equation}
\label{54}
\mathcal{V}_{Q_{x},Q_{y},Q_{z}} =\mathcal{V}_{Q_{x},Q_{z},Q_{y}}=\cdots=\mathcal{V}_{-Q_{x},Q_{y},Q_{z}} =\cdots
\end{equation}
which corresponds in real space to $\mathcal{V}(x,y,z)=\mathcal{V}(x,z,y)=\cdots=\mathcal{V}(-x,y,z)=\cdots$. We then find that, like for atoms, all the matrix elements in Eq.~(\ref{52}) are equal to zero except
 \begin{eqnarray}
\label{53}
 \lefteqn{\bra{z, v;\textbf{0}}     \mathcal{L}_y(\textbf{r})\ket{x, v;\textbf{0}}}\hspace{1.6cm}\nonumber\\
 \!\!
&{=}&\!\!i\hbar  \sum _{\textbf{K}'\textbf{K}} \mathcal{ V}_{\textbf{K}'-\textbf{K}}
 \mathcal{U}^*_{z, v;\textbf{K}'}   ( \textbf{K}'{\times} \textbf{K})_y     \mathcal{U}_{x, v;\textbf{K}}
  \nonumber   \\
 &\equiv &  i\hbar \,\,  \mathcal{W}\,,
\end{eqnarray}
and similar terms obtained from cyclic permutations.

  \subsection{On the state parity}
  
  Equation (\ref{53}) holds for even and odd states. To show it, let us consider $ \bra{\lambda' ,v;\textbf{0}} \mathcal{L}_x(\textbf{r})\ket{\lambda, v;\textbf{0}}$ for  three-fold states having an odd parity, that is, for $ \mathcal{U}_{\lambda, v;\textbf{K}}$ given in Eq.~(\ref{47}). This matrix element then appears as
  \begin{eqnarray}
\label{55}
 \bra{\lambda',v;\textbf{0}}     \mathcal{L}_x(\textbf{r})\ket{\lambda, v;\textbf{0}}
=i\hbar  \sum _{\textbf{K}',\textbf{K}}  G^*_o(K') G_o(K) \,\,\,\,\,\,\,\,\,\,\,\,\,\,\,\,\,\,\,\,
   \\
 \times K'_{\lambda'} K_{\lambda}
\mathcal{V}_{\textbf{K}'-\textbf{K}}
(K'_y K_z-K'_z K_y)\,. \nonumber
 \end{eqnarray}
 When $\lambda'=\lambda$, the above quantity is equal to zero for $\mathcal{V}_{Q_x,Q_y,Q_z}=\mathcal{V}_{Q_x,-Q_y,Q_z}$ (or $\mathcal{V}_{Q_x,Q_y,Q_z}=\mathcal{V}_{Q_x,Q_y,-Q_z}$), as seen by changing $K_y$ into $-K_y$ and $K'_y$ into $-K'_y$. When $\lambda'=x\neq \lambda$, it also is equal to zero for $\mathcal{V}_{Q_x,Q_y,Q_z}=\mathcal{V}_{-Q_x,Q_y,Q_z}$, as seen by changing $K_x$ into $-K_x$ and $K'_x$ into $-K'_x$. 
 
The same argument holds for three-fold states having  even parity, that is, for $ \mathcal{U}_{\lambda, v;\textbf{K}}$ given in Eq.~(\ref{48}). Note that linear combinations of even and odd valence states also give zero because the $\vec {\mathcal{L}}(\textbf{r})$ operator is an even operator, $\vec {\mathcal{L}}(-\textbf{r})=\vec {\mathcal{L}}(\textbf{r})$.

The last step is to diagonalize the spin-orbit interaction $H_{so}$ for even or odd valence states. This diagonalization follows exactly  the same procedure as the one for atoms in Sec.\ref{sec:3b}. Therefore, whatever their parity, the three-fold orbital states $(\lambda, v)$ with $(\pm1/2)$ spin split into two degenerate states and four degenerate states, these states having exactly the same structure as the ones for atoms given in Eq.~(\ref{23}). 

 \section{Discussion\label{sec:5}}
 
 \subsection{Spin-orbit splitting from group theory}

The group theory formalism and its tables of characters\cite{Koster,Falicov} commonly proposed to properly study the spin-orbit coupling in semiconductor crystals\cite{Cardona,Ivchenko} are, to our opinion, too cumbersome to deal with just three-fold orbital states. Still, in more complicated structures, using group theory may remain the only convenient way to tackle the problem. While this work purposely avoids using group theory, we nevertheless wish, for completeness, to recall some key results within this language.
 
Orbital states belong to the so-called ``simple group". The non-degenerate orbital state, which for atoms corresponds to $\ell=0$, is called $\Gamma_1$ in the case of semiconductor crystals, while the three-fold orbital state which for atoms corresponds to $\ell=1$, is called $\Gamma_5$.

\begin{figure}[!h]
\begin{center}
\includegraphics[width=8cm]{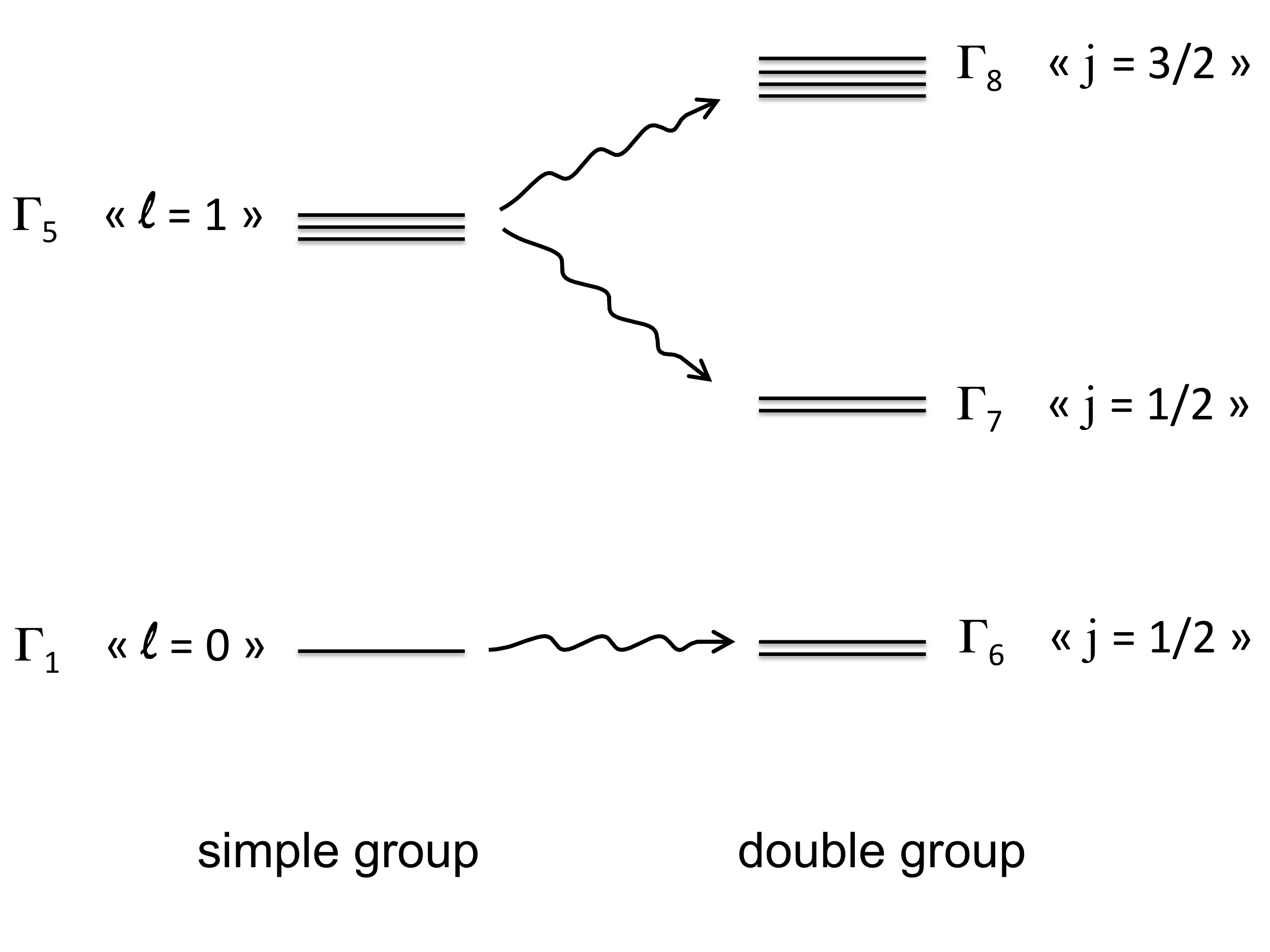}
\end{center}
\caption{Relevant semiconductor states according to the group theory irreducible representations and their denomination within atomic notations.  }
\label{fig:1}
\end{figure}

Including the spin degrees of freedom transforms the simple group into the ``double group". The $\Gamma_1$ state of the simple group gives rise to two states in the double group, called $\Gamma_6$, which for an atom would correspond to the two $j=1/2$ states (see Fig.~\ref{fig:1}). In the same way, the three $\Gamma_5$ states of the simple group, give rise to six states in the double group, which are further split by the spin-orbit interaction into four states called $\Gamma_8$, which correspond to  the four $j=3/2$ states of atoms, and into two states called $\Gamma_7$, which correspond to the two $j=1/2$ states of atoms. Note that despite the fact that they correspond to the same atomic quantum number $j=1/2$, the $\Gamma_6$ and $\Gamma_7$ states are fundamentally different because they are made of orbital states that respectively are non-degenerate and three-fold degenerate. Being eigenstates of the spin-orbit interaction with different eigenvalues, these states are orthogonal. 

\subsection{Physical understanding}

The fact that the eigenstates of the spin-orbit interaction for three-fold orbital states have the same structure regardless of the state parity, odd or even, and the potential symmetry, spherical or periodic, can be proven   in a mathematically rigorous way. For a spherical potential as in the case of atoms, the state symmetry is handled in the real space, while for a periodic potential as in the case of semiconductor crystals, it is handled  in the  reciprocal space. Similarity in these dual spaces is largely due to the fact that the spin-orbit eigenstates are derived for  electrons having two spin states only, $\ket{+1/2}$ and $\ket{-1/2}$, as seen from  Eq.~\eqref{28}. So, the spin can either stay the same or flip. For spin states quantized along the $z$ direction, the orbital operator associated with spin conservation is $\mathcal{L}_z$, while the orbital operator associated with  spin flip is $\mathcal{L}_{\eta}=\mathcal{L}_{x}+ i\eta \mathcal{L}_{y}$ with $\eta=\pm1$. In a bulk cubic crystal, the physically relevant orbital states for $\mathcal{L}_z$ and $\mathcal{L}_{\eta}$ are $\ket{0}=i\ket{z}$ and $\ket{\eta}=(- i\eta\ket{x}+\ket{y})/\sqrt{2}$, whatever the  degeneracy and symmetry of the electron states. 

Taking this key point into account, the remaining task is to find the set of orthogonal combinations of spin and orbital states, built on  $\ket{0}$ and $\ket{\pm1}$ that fulfills Eq.~\eqref{28}, whatever the symmetry of the potential felt by the electron. Within  group theory, orbital and spin states are mixed into the double group, which totally hides the state and potential symmetries. Indeed, the orbital state $\ket{\eta}$ can be associated with the  spin state $\ket{\eta/2}$ or $\ket{-\eta/2}$. Thanks to Eq.~\eqref{28}, we readily see that if $\ket{\eta}$ is associated with $\ket{-\eta/2}$, another orbital state $\ket{0}$ has to enter the eigenstate and this $\ket{0}$ state must have a $\ket{\eta/2}$ spin. This is easily seen from the $\textbf{J}$ eigenstates but the same argument stays valid for a periodic potential with cubic symmetry, that is, $(x, y, z)$ playing the same role. The proper combination of spin and orbital  states just follows from the spin-conserving and spin-flipping operators $\mathcal{L}_z$ and $\mathcal{L}_{\eta}$ appearing in Eq.~\eqref{28}.

\subsection{Two-dimensional materials}
The proposed procedure can be  extended to orbital states having a degeneracy higher than three-fold, like for materials, that have recently attracted a lot of interest. In single-layer graphene\cite{graphene} and transition metal dichalcogenides\cite{Liu}, the $d$ orbital states appear to play a more important role in the spin-orbit splitting than the $p$ orbital states. 

For single-layer graphene,  the $D_{3h}$ crystal symmetry at the K and K$'$ points allow the two higher $d_{\pm1}$ orbital states to enter into play in the $\pi$ band\cite{graphene}; so, the relevant orbital states of the problem at the Dirac points are $|0\rangle=i|z\rangle$ and $|d_{\pm1}\rangle=(\mp i|xz\rangle+|yz\rangle)/\sqrt{2}$. The same equation \eqref{28} leads us to see that if $\ket{d_\eta}$ is associated with $\ket{-\eta/2}$, the other spin state $\ket{\eta/2}$ that enters the eigenstate must have an orbital state  $\ket{d_0}$, which is absent at the Dirac points. As a result, the spin-orbit splitting comes from the spin-conserving operator $\mathcal{L}_z$ between $\ket{d_\eta}$,  in addition to small second-order contribution from the $|0\rangle$ state of the $\pi$ band and the $|\eta\rangle$ states of the $\sigma$ band.

  For transition metal dichalcogenides, the situation is even more complex because all five $d$ orbital states from the metal atom and the $p$ orbital states from the chalcogen atom  play a role in the spin-orbit splitting. The study of these complex materials is beyond the scope of the present work.  Yet, in view of its simplicity for the $p$ orbital states, we expect the present procedure  to be quite valuable to physically understand the spin-orbit eigenstates of these complex structures.

\subsection{Misleading notations}
We would like to end this work by stressing that not only it is physically incorrect to extend the spin-orbit procedure for atoms to periodic crystals but, far worse, labeling the spin-orbit eigenstates in the same way, $(j, j_z)$, as if electrons in a crystal had an orbital momentum, is quite misleading. In particular, this tends to mess up the whole understanding of the exciton-photon interaction. 

Indeed, what is commonly said is the following: valence holes come from $p$ valence states; they thus have an orbital momentum $\ell=1$, which with their spin $\textsl{s}=1/2$, gives them a total momentum $\textbf{J}=\textbf{L}+\textbf{S}$ with $j^h=(3/2,1/2)$. Conduction electrons are also said to be in a $s$ state, with an orbital momentum $\ell=0$; so, they are only labeled by their spin $\textsl{s}_{z}=\pm1/2$. This would give to the conduction electron-valence hole pairs a total momentum $\textbf{J}^{eh}=\textbf{S}^e+\textbf{J}^h$, with $j^{eh}=(2,1,0)$. It is then claimed that the two excitons $(+2,-2)$ made from  electron-hole pairs $(j^{eh}=2,j^{eh}_z=\pm2)$ are dark because they cannot be coupled to photons since photons have a ``spin'' $(\pm1,0)$ which correspond to polarizations $\sigma_\pm$ and $\pi$.\

\begin{figure}[!t]
\begin{center}
\includegraphics[width=8cm]{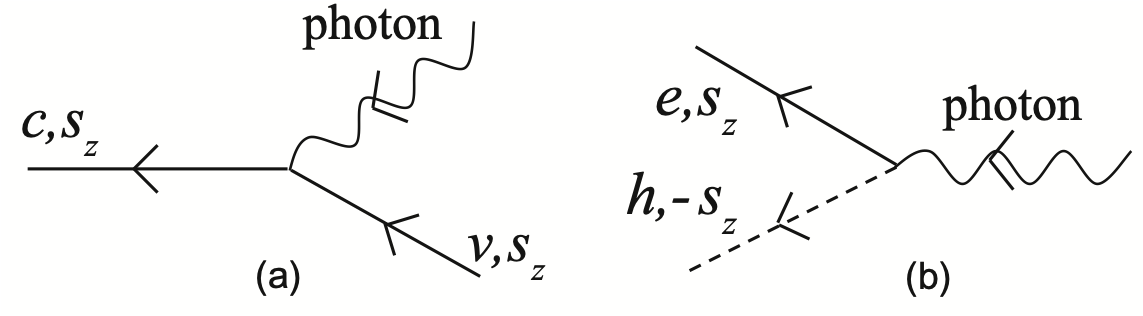}
\end{center}
\caption{(a) Absorption of a photon excites an electron  from the valence band \textit{v} to the conduction band $c$ while keeping its electron spin $\textsl{s}_{z}$. (b) In terms of electron and hole, the absorption of a photon creates an electron-hole pair with zero total spin. }
\label{fig:2}
\end{figure}
The correct understanding is quite different, even if in the very end only two exciton states are not coupled to photons, these excitons having the lowest energy for the very same reason that they are dark.

$\bullet$ First, photons are known not to have spin in the proper sense but a two-fold polarization associated with a vector in the two-dimensional plane perpendicular to the photon propagation axis $\mathbf{e}_Z=\bQ/Q$. This polarization vector is a linear combination of $(\mathbf{e}_X,\mathbf{e}_Y)$, the one associated with circular polarizations being 
\begin{equation}
\mathbf{e}_{\pm1}=\frac{\mp i \mathbf{e}_X+\mathbf{e}_Y}{\sqrt{2}}\,.
\end{equation}
In an electron-photon interaction, the photon does not act on the electron spin but, via its polarization, it induces a change in the electron orbital wave function. So, the fact that the photon ``spin'' is never equal to $2$, cannot be linked to the fact that excitons, commonly labeled as $2$ or $-2$, are dark.

$\bullet$ Actually, it is just because the photon does not change the electron spin that some exciton states are dark, the fact that there are two dark states only being a direct consequence of the spin-orbit interaction.

When going from the valence band to the conduction band under a photon absorption, the  electron conserves its spin. As the hole spin is opposite to the spin of the missing valence electron, the electron-hole pair coupled to photon through its absorption, has a total spin equal to zero (see Fig.~\ref{fig:2}).

In the absence of spin-orbit interaction, there are $3\times 2=6$ hole states, labeled by $\lambda=(x,y,z)$ and $\textsl{s}_{z}=\pm1/2$ and the two electron states labeled by their spin only, for a non-degenerate conduction band; so, they would be six excitons with total spin equal to zero, namely $\lambda=(x,y,z)$ and $\textsl{s}^e_z=-\textsl{s}^h_z=\pm1/2$.

We have shown that the spin-orbit interaction couples the spin and orbital indices of the valence electron in such a way that there are two states  only with a well-defined spin, namely $|\eta\rangle\otimes|\eta/2\rangle$, the other four valence states being linear combinations of states with $1/2$ and $-1/2$ spins. The destruction of such valence electrons $|\eta\rangle\otimes|\eta/2\rangle$ which are pure in spin, leads to hole states also pure in spin. So, when combined with a conduction electron having a $\eta/2$ spin, they form two dark excitons with total spin 1 and two bright excitons with total spin 0.

As a result, the existence of two dark exciton states \textit{only} is due to the fact that: (i) the electron-photon interaction conserves the spin, and (ii) two valence states only stay pure of spin under the spin-orbit coupling.

$\bullet$ All this shows that inadequate notations tend to lead to incorrect physical understanding. Even if done for years, we suggest to stop labeling spin-orbit valence states by $(j,j_z)$ as for atomic states, but by $(\zeta,\zeta_z)$ and call them \textit{valence spin-orbit indices}. This would prevent considering $\zeta$ as a naive angular momentum and adding it to the conduction spin, to end with a ``total angular momentum of the exciton", which is physically meaningless.

\begin{figure}[!t]
\begin{center}
\includegraphics[width=8cm]{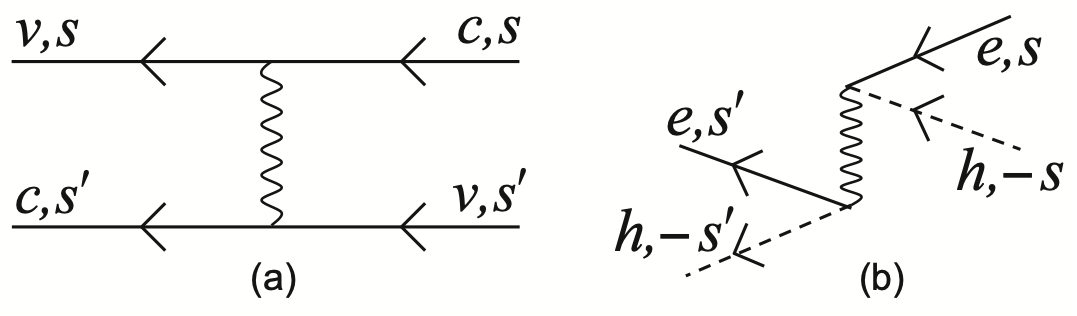}
\end{center}
\caption{ (a) Interband Coulomb interaction conserves the spin $\textsl{s}$. (b) Electron-hole pairs having a zero total spin can couple to photon. }
\label{fig:3}
\end{figure}

$\bullet$ Finally, Coulomb interaction conserves spin, just as electron-photon interaction does. So, the electron-hole pairs that suffer interband Coulomb interaction have a total spin equal to zero, just like the pairs that are coupled to photons (see Fig.~\ref{fig:3}). This (repulsive) Coulomb interaction pushes the energy of bright excitons above the one of dark excitons. So, the excitons that have the lowest energy, are dark for the very same reason that they are not coupled to photons.

 \section{Conclusion}

This work considers a very fundamental aspect of semiconductor crystals that is either ignored when treating the valence band as a true $p$ state, or not physically understood when using  group theory. 

 We present a direct procedure---easy to follow by anyone with no  background on group theory---to derive the spin-orbit energy shifts of three-fold orbital states in cubic semiconductor crystals, whatever the parity of these states. 
We show that the state degeneracy matters, but not the state parity, even or odd, nor the potential symmetry, spherical or periodic.

We show that the spin-orbit eigenstates have the same structure for semiconductor crystals and atoms with same orbital degeneracy, and we physically explain why this has to be so. Nevertheless, we urge to stop calling these valence states through ($j,j_z$) indices like atoms, and mostly to stop relating  these indices to ``total angular momentum" because $\textbf{L}$ and $\textbf{J}$ only have a meaning for problem with spherical symmetry.

The simplicity of the approach we here propose to spin-orbit interaction should appear as quite valuable in the case of complex materials having valence electrons with degeneracy higher than three-fold, as many materials of today major interest. We leave these studies to future works.

\subsection{Acknowledgments}
The authors would like to thank Yia-Chung CHANG for  enlightening arguments on the parity of valence and conduction states, and also Benoit EBLE for discussions at the beginning of this work.

\appendix
%\numberwithin{equation}{section}

\section{Spin-orbit interaction\label{app:A}} 

The spin-orbit interaction is most often described in the context of atomic physics. This is why it is commonly identified with the $\textbf{L}\cdot\textbf{S}$ interaction. To get the whole story straight is quite complicated, in particular the derivation of the $1/2$ factor which enters the prefactor of the coupling. The derivation we here  present  is a combination of what can be found in various textbooks, in particular Baym \cite{Baym}, Tomonaga\cite{Tomonaga}, and Landau-Lifschitz\cite{LL}, with the  aim to make the presentation as simple as possible,  a particular attention being paid  on choosing transparent notations. 

\subsection{Magnetic moments}
We consider an electron with mass $m_0$ and charge $e= -|e|$.\\
$\bullet$ The electron angular momentum reads as 
  \begin{equation}
\label{A1}
\textbf{r}\times \textbf{p}= \textbf{L} = \hbar \vec{\ell}\,,
\end{equation} 
 where $\textbf{p}=(\hbar/i)\,\overrightarrow{\nabla}$ is the electron kinetic momentum,  and $|\ell |$ is an integer, the $ \vec{\ell}$ projection $\ell_z$ on the $z$ axis taking any integer values between $(|\ell |, -|\ell |)$. The electron magnetic  moment associated with its orbital angular momentum is 
 \begin{equation}
\label{A2}
 \textbf{M}_{L}=\frac{e}{2m_{0}c} \textbf{L} \equiv \mu_{B} \vec{\ell}\,,
\end{equation}
where $\mu_{B} =|e|\hbar/2m_{0}c$ is called Bohr magneton. Its value is $9.274\times 10^{-21}$ erg/Gauss.\\
$\bullet$  The electron also has a magnetic moment associated with its spin $\textbf{S}=(\hbar/2)  \, \vec{\sigma} $ where the components of the $\vec{\sigma}$  vector are the Pauli matrices $(\sigma_x,\sigma_y,\sigma_z)$. It reads
\begin{equation}
\label{A3}
\textbf{M}_{S}= g_{e}\frac{e}{2m_{0}c}\textbf{S}\,,
\end{equation}
the Land\'e factor for electron being $g_{e}=2$.\\
$\bullet$ So, the total magnetic moment of an electron with  orbital angular momentum $\textbf{L}$ and  spin $\textbf{S}$ is given by
\begin{equation}
\label{A4}
\textbf{M}=\textbf{M}_{L}+\textbf{M}_{S}=\frac{e}{2m_{0}c} \big(\textbf{L}+g_{e}\textbf{S}\big)\,.
\end{equation}
$\bullet$ On the other hand, in an external magnetic field $\textbf{H}_{ext}$, that is, a field which has nothing to do with the charge and velocity of the particle at hand, this particle has a magnetic energy associated with the magnetic moment $\textbf{M}$  equal to
\begin{equation}
\label{A5}
\mathcal{E}_{mag}=-\textbf{M}\cdot \textbf{H}_{ext}\,.
\end{equation}

\subsection{Magnetic field}
$\bullet$  Let us consider an electron moving with a velocity $\textbf{v}=\textbf{p}/m_0$ in an electrostatic potential $\Phi_{ext}(\textbf{r})$ due to an external electric field $\textbf{E}_{ext}(\textbf{r})$. Its Hamiltonian reads 
  \begin{equation}
\label{A6}
\mathcal{H}_{0}=\frac{\textbf{p}^2}{2m_{0}}+e \Phi_{ext}(\textbf{r})\,.
\end{equation}
We put this electron in a strong external magnetic field $\textbf{H}_{ext}$. (We will see later on why this field has to be strong for the following to be true). The vector potential associated with $\textbf{H}_{ext}~=~\textbf{rot}\, \textbf{A}_{ext}$ reads, in the Coulomb gauge, as
\begin{equation}
\label{A7}
\textbf{A}_{ext}=\frac{1}{2}\textbf{H}_{ext}\times \textbf{r}\,,
\end{equation}
 which indeed fulfills $\mathrm{div}\textbf{A}_{ext}=0$. If we neglect  spin, the electron Hamiltonian would be 
\begin{equation}
\label{A8}
\frac{1}{2m_{0}}\big(\textbf{p} -\frac{e}{c} \textbf{A}_{ext}\big)^{2}+ e\Phi_{ext}\,.
\end{equation}

 The spin magnetic moment brings an additional energy which, according to Eq. \eqref{A5}, reads 
 \begin{equation}
\label{A9}
-\textbf{M}_{S}\cdot \textbf{H}_{ext}=-g_{e}\frac{e}{2m_{0}c} \textbf{S}\cdot \textbf{H}_{ext}\,.
\end{equation}
 So, with this spin contribution, the electron Hamiltonian  appears as
 \begin{equation}
\label{A10}
\mathcal{H}=\frac{1}{2m_{0}}\big(\textbf{p} -\frac{e}{c} \textbf{A}_{ext}\big)^{2}-g_{e}\frac{e}{2m_{0}c} \textbf{S}\cdot \textbf{H}_{ext}+ e\Phi_{ext}\,.
\end{equation}
By noting that $\textbf{A}_{ext}\cdot \textbf{p}=\textbf{p}\cdot \textbf{A}_{ext}$, as fulfilled by $\textbf{A}_{ext}$ given in Eq.~\eqref{A7}, we can rewrite the term linear in $\textbf{A}_{ext}$ as
\begin{eqnarray}
\label{A11}
 -\frac{e}{m_{0}c}\textbf{A}_{ext}\cdot \textbf{p} & = & \frac{-e}{2m_{0}c}(\textbf{H}_{ext}\times \textbf{r})\cdot \textbf{p} \nonumber\\
 &= &  \frac{-e}{2m_{0}c}(\textbf{r}\times \textbf{p})\cdot \textbf{H}_{ext}\,.
\end{eqnarray}
The Hamiltonian $\mathcal{H}$ given in Eq.~\eqref{A10} then appears at first order in $\textbf{H}_{ext}$ as
\begin{equation}
\label{A12}
\mathcal{H} = \mathcal{H}_{0}-\frac{e}{2m_{0}c} \big(\textbf{L}+g_{e}\textbf{S}\big)\cdot \textbf{H}_{ext}\,.
\end{equation}
The second term is just the magnetic energy given in Eq.~\eqref{A5} associated with the electron magnetic moment given in Eq.~\eqref{A4}.

$\bullet$ The above Hamiltonian is valid when the external field $\textbf{H}_{ext}$ is large compared to the internal field felt by the moving electron. This internal field follows from the fact that, due to the Lorentz transformation, the electromagnetic fields $(\textbf{E}, \textbf{H})$ become $(\textbf{E}', \textbf{H}')$ in a frame moving at a constant velocity $\textbf{v}$. The link between $(\textbf{E}, \textbf{H})$ and $(\textbf{E}', \textbf{H}')$ reads 
\begin{equation}\label{A13}
\begin{aligned}
\textbf{H}_{\|}' & = \textbf{H}_{\|}\,,     &  \textbf{E}_{\|}'& = \textbf{E}_{\|}\,,  \\
\textbf{H}_{\bot}' &= \frac{\textbf{H}_{\bot}{-}\frac{\textbf{v}}{c}\times \textbf{E}}{\sqrt{1-v^{2}/c^{2}}}\,, &\textbf{E}_{\bot}' &= \frac{\textbf{E}_{\bot}{+}\frac{\textbf{v}}{c}\times \textbf{H}}{\sqrt{1-v^{2}/c^{2}}}\,.
\end{aligned}
\end{equation}
So, the field components $(\textbf{H}_{\|}, \textbf{E}_{\|})$  parallel to $\textbf{v}$ do not change under the Lorentz transformation, but the field components $(\textbf{H}_{\bot}, \textbf{E}_{\bot})$  perpendicular to $\textbf{v}$ do. These equations are commonly referred to as Biot-Savart law.
Consequently, using Eqs.~(\ref{A3}, \ref{A13}), an electron with velocity $\textbf{v}$ feels an additional magnetic energy given by 
\begin{equation}
\label{A14}
 -g_{e}\frac{e}{2m_{0}c}\textbf{S}\cdot \Bigg{(}\frac{\frac{-\textbf{v}}{c}\times \textbf{E}_{ext}}{\sqrt{1-v^{2}/c^{2}}}\Bigg{)}\simeq - g_{e}\frac{e}{2m_{0}c} \textbf{S} \cdot \frac{\textbf{E}_{ext}\times \textbf{v}}{c}\,.
\end{equation}

$\bullet$ Actually, this is not fully correct because the electron in an electrostatic field $\textbf{E}'$ feels a force $e\textbf{E}'$ that produces an acceleration; so, the electron velocity  changes. As a result, the Lorentz transformation \eqref{A13},  valid for a  frame moving at constant velocity, cannot correctly give the internal magnetic field felt by the accelerating electron. As first shown by Thomas\cite{Thomas} and confirmed by Dirac\cite{Dirac},  the changing velocity has the effect of changing $g_e$ into $(g_{e}-1)$ in Eq.~\eqref{A14}. Therefore, the spin contribution to the energy of a spin-$\textbf{S}$ electron having a velocity $\textbf{v}$ in an external magnetic field $\textbf{H}_{ext}$ and  electrostatic field $\textbf{E}_{ext}$, ultimately reads as 
\begin{equation}
\label{A.15}
- \textbf{S}\cdot \left[g_{e}\frac{e}{2m_{0}c} \textbf{H}_{ext}+(g_{e}-1)\frac{e}{2m_{0}c}\frac{\textbf{E}_{ext}\times \textbf{v}}{c}\right]\,.
\end{equation}

\subsection{Pauli and Dirac equations}
$\bullet$ The Dirac equation for an electron in an external magnetic field $\textbf{H}_{ext}~=~\textbf{rot} \,\textbf{A}_{ext}$ reduces, up to  terms in $1/c$, to the Pauli equation\cite{Pauli}, namely
\begin{multline}
\label{A16}
 i\hbar \frac{\partial \varphi}{\partial t}  = \\
 \bigg( m_{0}c^{2}+\frac{1}{2m_{0}}\big(\textbf{p}-\frac{e}{c}\textbf{A}_{ext}\big)^{2}+ e\Phi_{ext} 
 -\frac{e\hbar}{2m_{0}c} \vec{\sigma} \cdot \textbf{H}_{ext}\bigg)\varphi \,.
\end{multline}
By writing the spin term as
\begin{equation}
\label{A17}
-\frac{e}{m_{0}c}\frac{\hbar}{2} \vec{\sigma}\cdot \textbf{H}_{ext}=-\frac{g_{e}}{2}\frac{e\hbar}{m_{0}c} \textbf{S}\cdot \textbf{H}_{ext}\,,
\end{equation}
we readily see that the electron Land\'e factor, $ g_{e}=2$,  implicitly appears in this equation. 

$\bullet$ If we go one step further and write the Dirac equation\cite{LL} up to terms in $1/c^2$, we find
\begin{eqnarray}
\label{A18}
i \hbar  \frac{\partial \Psi}{\partial t} & = & \left[ m_{0}c^{2}+\frac{1}{2m_{0}}\big(\textbf{p}-\frac{e}{c}\textbf{A}_{ext}\big)^{2}-\frac{p^{4}}{8m_{0}^{3}c^{2}}\right] \Psi \nonumber\\
 & - & \left[\frac{e\hbar}{2m_{0}c} \vec{\sigma} \cdot \textbf{H}_{ext} +\frac{e\hbar}{4m_{0}^{2}c^{2}} \vec{\sigma} \cdot \textbf{E}_{ext}\times \textbf{p}\right]\Psi \nonumber\\
 &+& \left[e\Phi_{ext} +\frac{\hbar^{2}e}{8m_{0}^{2}c^{2}}\mathrm{\Delta}\Phi_{ext}\right]\Psi\,,
\end{eqnarray}
with $\mathrm{\Delta}$ being the Laplace operator.

\textbf{(i)} The first bracket follows from the $1/c$ expansion of
\begin{equation}
\label{A19}
c\sqrt{m_{0}^{2}c^{2}+p^{2}}\simeq m_{0}c^{2}+\frac{p^{2}}{2m_{0}}-\frac{p^{4}}{8m_{0}^{3}c^{2}}\,.
\end{equation}

\textbf{(ii)} The second bracket of Eq.~\eqref{A18} corresponds to the spin contribution. The part coming from the electrostatic field $\textbf{E}_{ext}$ can be rewritten as
 \begin{equation}
\label{A20}
-\frac{e}{2m_{0}c}\frac{\hbar \vec{\sigma}}{2}\cdot\frac{\textbf{E}_{ext}\times \frac{\textbf{p}}{m_{0}}}{c}=- (g_{e}-1)\frac{e}{2m_{0}c}\textbf{S} \cdot \frac{\textbf{E}_{ext}\times \textbf{v}}{c}\,,
\end{equation}
since $g_{e}-1=1$. This $1/c^{2}$ term leads to the spin-orbit interaction, absent in the Pauli equation given in Eq.~\eqref{A16}. So, the total spin  contribution to the Dirac equation up to $1/c^{2}$ terms  reads as
\begin{equation}
\label{A21}
- g_{e}\frac{e}{2m_{0}c} \textbf{S} \cdot \Bigg(\textbf{H}_{ext}+ \frac{\textbf{E}_{ext}\times \textbf{v}}{c}\Bigg)+\frac{e}{2m_{0}c} \textbf{S} \cdot \frac{\textbf{E}_{ext}\times \textbf{v}}{c}\,.
\end{equation} 
The first term corresponds to the  fields seen in a frame having a constant velocity $\textbf{v}$, which are Lorentz-transformed from the $(\textbf{H}_{ext}, \textbf{E}_{ext})$ fields of the laboratory frame. The second term corresponds to the Thomas' correction due to the electron acceleration induced by the electrostatic force $e\textbf{E}_{ext}$. This force leads to effectively  replacing $g_e$ with $(g_{e}-1)$ but in the $\textbf{S} \cdot (\textbf{E}_{ext}\times \textbf{v})$ term only.

\textbf{(iii)} The second term in the last bracket of Eq.~\eqref{A18} differs from zero when local charges are present, as can be seen from the Ohm's law, $\mathrm{div}\Phi_{ext}(r)=-4\pi \rho_{ext}(r)$ where $\rho_{ext}(r)$ is the charge density.

\subsection{Thomas' understanding\label{app:Thomas}}
The Dirac equation definitely gives the correct  spin-orbit interaction, including its numerical prefactor. However, it is hard from it to physically catch why the Land\'e factor $g_e$ is changed to $(g_{e}-1)$ in one part only of the magnetic energy. 

 The derivation of the change from $g_e$ to $(g_{e}-1)$ proposed by Thomas has the great advantage to trace its physics to the fact that the electron velocity is not constant, due to the presence of the electrostatic field $\textbf{E}_{ext}$ and the force this field induces on the electron. Thomas' result is identical to the one obtained from the Dirac equation taken up to terms in $1/c^2$. Actually, his result pushed Pauli to admit that relativistic quantum theory is not the only way to handle the spin properly. 

\begin{figure}[!h]
\begin{center}
\includegraphics[width=8cm]{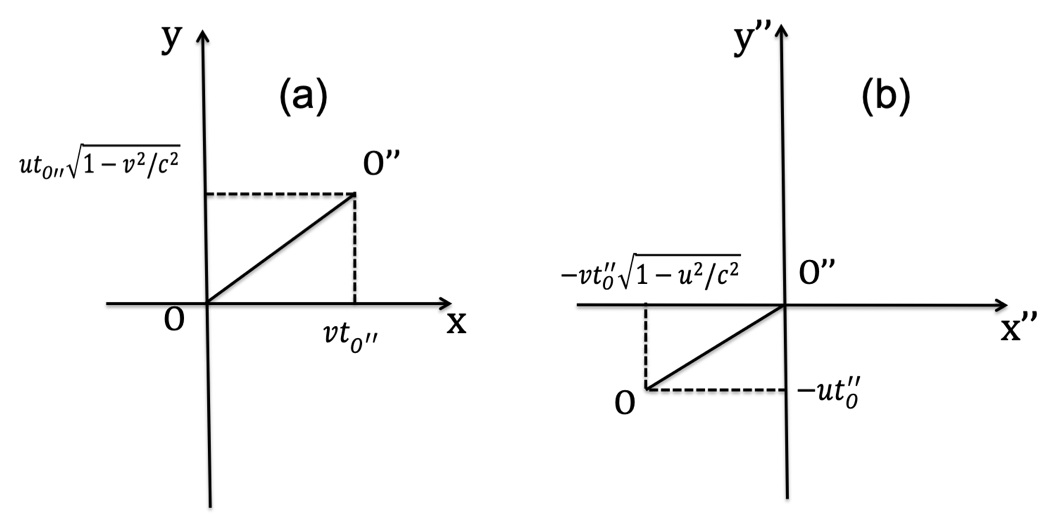}
\end{center}
\caption{Components of the $\mathrm{O}''$ velocity in the F frame (a) and the O velocity in the frame $\mathrm{F}''$ (b).}
\label{fig:OOab}
\end{figure}

Thomas' derivation can be divided into four steps:\\
\textbf{(1)} First, we introduce a frame F that we call laboratory frame, a frame $\mathrm{F}'$ moving with a velocity $\textbf{v}$ 
along $\textbf{x}$, and another frame $\mathrm{F}''$ moving with a velocity $(\textbf{v}+\textbf{u})$, with $\textbf{u}$ along $\textbf{y}$. \\
\indent The coordinates $(x,y,z,t)$  in the F frame and $(x', y', z', t')$ in the $\mathrm{F}'$ frame are related by a Lorentz transformation, namely
\begin{equation}
\label{A23}
x'=\frac{x-vt}{\sqrt{1-v^2/c^2}}, \quad y'=y, \quad z'=z ,\quad t'=\frac{t-\frac{v}{c^2}x}{\sqrt{1-v^2/c^2}}.
\end{equation}
As the $\mathrm{F}''$ frame has a $\textbf{u}$ velocity with respect to the $\mathrm{F}'$ frame, the coordinates $(x'', y'', z'', t'')$ in the $\mathrm{F}''$ frame and $(x', y', z', t')$ in the $\mathrm{F}'$ frame are also related by a Lorentz transformation. So, $(x'', y'', z'', t'')$ in the $\mathrm{F}''$ frame  read in terms of $(x,y,z,t)$ in the F frame  as
\begin{equation}
\begin{aligned}\label{A24}
x'' & =   x'= \frac{x-vt}{\sqrt{1-v^2/c^2}} ,  \\
y''  & =   \frac{y'-ut'}{\sqrt{1-u^2/c^2}}=\frac{y\sqrt{1-v^2/c^2}+\frac{uv}{c^2}x-ut}{\sqrt{(1-u^2/c^2)(1-v^2/c^2)}} ,\\
z'' & =  z'= z \\
t''  & =   \frac{t'-\frac{u}{c^2}y'}{\sqrt{1-u^2/c^2}}=\frac{t-\frac{v}{c^2}x-\frac{u}{c^2}\sqrt{1-v^2/c^2}\,y}{\sqrt{(1-u^2/c^2)(1-v^2/c^2)}}
\end{aligned}
\end{equation}
\textbf{(2)} We now consider the $\mathrm{O}''$ origin of the $\mathrm{F}''$ frame. Its coordinates in the $\mathrm{F}''$ frame are by construction $0=x''_{_{\mathrm{O}''}}=y''_{_{\mathrm{O}''}}=z''_{_{\mathrm{O}''}}$ whatever $t''_{_{\mathrm{O}''}}$.
Equation \eqref{A24} gives its coordinates ($x_{_{\mathrm{O}''}},y_{_{\mathrm{O}''}},z_{_{\mathrm{O}''}}$) in the F frame through
\begin{equation}
\begin{aligned}
\label{A25}
 0 & =  x_{_{\mathrm{O}''}}- vt_{_{\mathrm{O}''}}\, \\
 0 & =   y_{_{\mathrm{O}''}} \sqrt{1-v^2/c^2}+ \frac{uv}{c^2}x_{_{\mathrm{O}''}}-ut_{_{\mathrm{O}''}} \, \\
 0 & =   z_{_{\mathrm{O}''}}\,
\end{aligned}
\end{equation}
which leads to
\begin{equation}
\begin{aligned}
\label{A26}
x_{_{\mathrm{O}''}}& =   vt_{_{\mathrm{O}''}} \, \\
y_{_{\mathrm{O}''}} &  =   \frac{ut_{_{\mathrm{O}''}}-\frac{uv}{c^2}x_{_{\mathrm{O}''}}}{\sqrt{1-v^2/c^2}}=ut_{_{\mathrm{O}''}}\sqrt{1-v^2/c^2} \, \\
z_{_{\mathrm{O}''}} & =  0 \,
\end{aligned}
\end{equation}
So, the components of the velocity $\textbf{v}_{_{\mathrm{O}''}}$ of the $\mathrm{F}''$ frame origin, $\mathrm{O}''$, are in the F frame not equal to $(v, u, 0)$ but to $(v, u\sqrt{1-v^2/c^2}, 0)$ (see Fig.~\ref{fig:OOab}a). \\

\indent In the same way, the coordinates of the O origin of the F frame are $0=x_{_{\mathrm{O}}}=y_{_{\mathrm{O}}}=z_{_{\mathrm{O}}}$ whatever $t_{_{\mathrm{O}}}$. Equation \eqref{A24} gives them in the $\mathrm{F}''$ frame through 
\begin{equation}
\begin{aligned}
\label{A27}
x''_{_{\mathrm{O}}} & =  -\frac{vt_{_{\mathrm{O}}}}{\sqrt{1-v^2/c^2}}=-vt_{_{\mathrm{O}}}''\sqrt{1-u^2/c^2}\,,\\
y''_{_{\mathrm{O}}} & =   -\frac{ut_{_{\mathrm{O}}}}{\sqrt{(1-v^2/c^2)(1-u^2/c^2)}}=-ut_{_{\mathrm{O}}}'' \,, \\
z''_{_{\mathrm{O}}} & =  0  \,, \\
t''_{_{\mathrm{O}}} & =  \frac{t_{_{\mathrm{O}}}}{\sqrt{(1-v^2/c^2)(1-u^2/c^2)}}\,.
\end{aligned}
\end{equation}
From these equations, we find that the components of the velocity $\textbf{v}''_{_{\mathrm{O}}}$ of the F frame origin, O, are in the $\mathrm{F}''$ frame not equal to ($-v,-u,0$) but to $(-v\sqrt{1-u^2/c^2}, -u,0)$ (see Fig.~\ref{fig:OOab}b).\\

\begin{figure}[!h]
\begin{center}
\includegraphics[width=8cm]{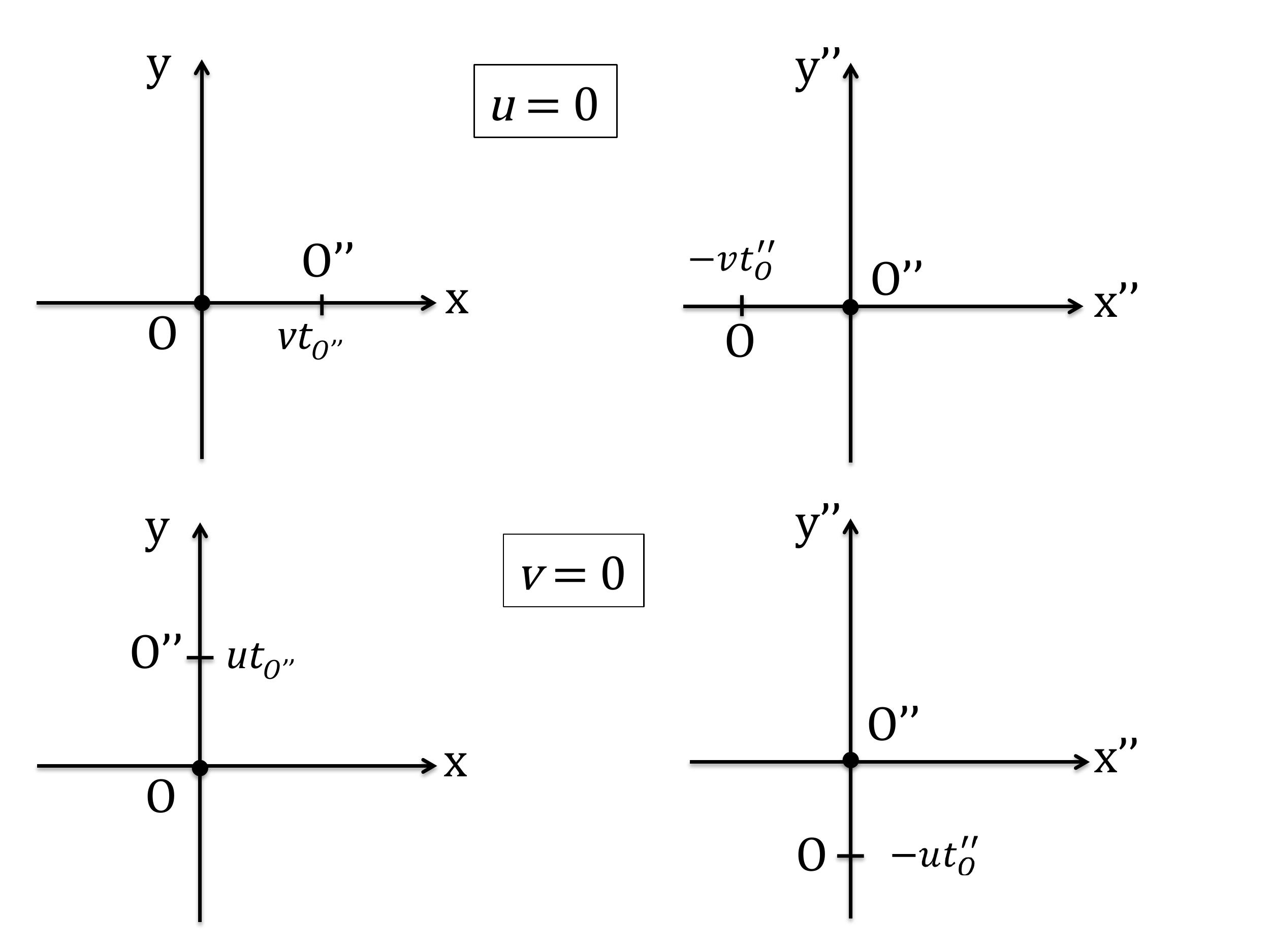}
\end{center}
\caption{Motion of the (F, $\mathrm{F}''$) frames when $uv=0$.  }
\label{fig:uv0}
\end{figure}

\begin{figure}[!h]
\begin{center}
\includegraphics[width=8cm]{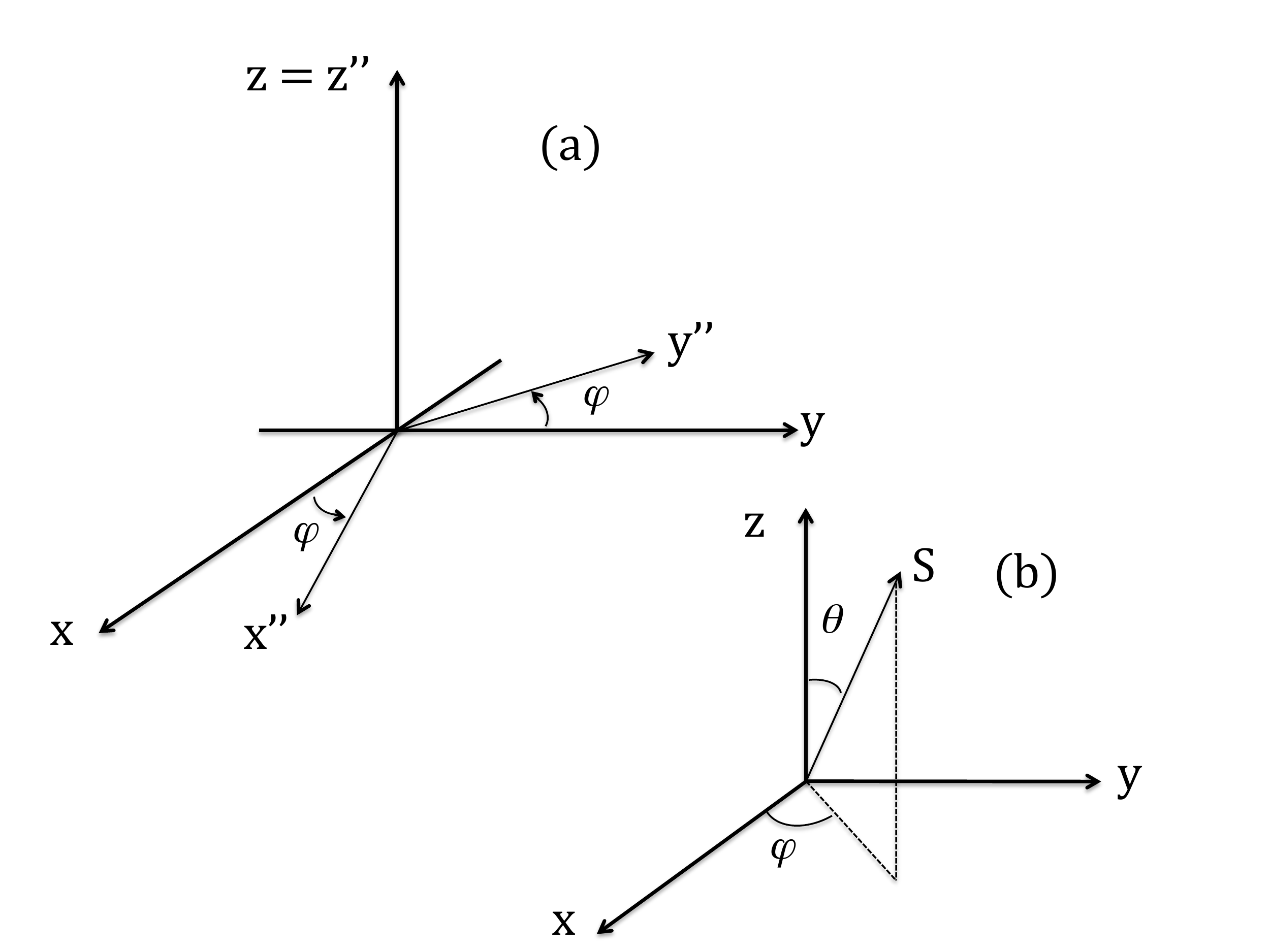}
\end{center}
\caption{(a) The frame $\mathrm{F}''$  and (b) the frame F are related by a rotation around the $z$ axis when $uv\neq0$.}
\label{fig:anglerot}
\end{figure}

\indent Before going further, let us discuss these results. The two velocities $\textbf{v}_{_{\mathrm{O}''}}$ and $\textbf{v}''_{_{\mathrm{O}}}$ have the same modulus $\sqrt{u^2+v^2-u^2v^2/c^2}$. This value differs from the modulus of the velocity ($\textbf{u}+\textbf{v}$) of the $\mathrm{F}''$ frame with respect to the F frame when $uv\neq0$.  Indeed, when  $u$ or $v$ is equal to zero, $\textbf{v}''_{_\mathrm{O}}=-\textbf{v}_{_{\mathrm{O}''}}$ and the F and $\mathrm{F}''$ frames are related by a bare translation, as seen from Fig.~\ref{fig:uv0}. By contrast, when $uv\neq0$, the two frames are related by a rotation around the $z$ axis. To obtain this rotation, we note that $(x'', y'')$ obtained from $(x, y)$ by a $\varphi$ rotation around $z=z''$, are given by (see Fig.~\ref{fig:anglerot})
\begin{equation}
\begin{aligned}
\label{A28}
x'' & =  x\cos\varphi+y\sin\varphi \,, \\
y'' & =  y\cos\varphi-x\sin\varphi \,.
\end{aligned}
\end{equation}

When used for the $\textbf{v}''_{_\mathrm{O}}$ velocity in the $\mathrm{F}''$ frame and the $\textbf{v}_{_{\mathrm{O}''}}$ velocity in the F frame, that is, for $(-v\sqrt{1-u^{2}/c^{2}},-u)$ and $(v, u\sqrt{1-v^{2}/c^{2}})$ (see Fig. 4), these two equations give, since $-\textbf{v}''_{_\mathrm{O}}=\textbf{v}_{_{\mathrm{O}''}}$ for $\varphi =0$,
\begin{equation}
\begin{aligned}
\label{A29}
 -(-v\sqrt{1{-}u^2/c^2})  &= v\cos\varphi +u\sqrt{1{-}v^2/c^2}\sin\varphi \,, \\
 -(-u) & =  u\sqrt{1{-}v^2/c^2}\cos\varphi - v\sin\varphi \,.
\end{aligned}
\end{equation}
So, the rotation angle $\varphi$ of the $\mathrm{F}''$ frame with respect to the F frame is related to $(u,v)$ through
\begin{eqnarray}
\cos\varphi &=&  \frac{v^2\sqrt{1-u^2/c^2}+u^2\sqrt{1-v^2/c^2}}{u^2+v^2-u^2v^2/c^2}\,,  \label{A30ab}
\\
\sin\varphi  &=&  uv \frac{\sqrt{(1-v^2/c^2)(1-u^2/c^2)}-1}{u^2+v^2-u^2v^2/c^2}\,.\label{A31}
\end{eqnarray}
For $uv=0$, the above equations give $\varphi=0$, that is no rotation, as expected.

\textbf{(3)} Next, we consider the origin $\mathrm{O}''$ of the $\mathrm{F}''$ frame  in the F frame. When $t=0$, the $\mathrm{O}''$ point  has a velocity $\textbf{v}$ along $\textbf{x}$. After a short time delay $\Delta t$, its velocity is $(\textbf{v}+\textbf{u})$, which corresponds to a velocity change $\Delta \textbf{v}= \textbf{u}$ along $\textbf{y}$.
The $(x_{_{\mathrm{O}''}}, y_{_{\mathrm{O}''}}, z_{_{\mathrm{O}''}})$ coordinates of $\mathrm{O}''$ in the F frame are then given, according to Eq.~\eqref{A26} and as illustrated in Fig.~\ref{fig:OOab}a, by
\begin{equation}
\label{A32}
(v\Delta t\,, \Delta v\sqrt{1-v^2/c^2}\Delta t\,, 0)
\end{equation}
When compared to the velocity $(v,0,0)$ in the F frame for $t=0$, the velocity $(v, \Delta v\sqrt{1-v^2/c^2},0)$ after a time delay $\Delta t$ corresponds to  an acceleration 
\begin{equation}
\label{A32a}
a=\sqrt{1-v^2/c^2} \,\,  \frac{\Delta v}{\Delta t}
\end{equation}
along $\textbf{y}$. This acceleration brings a rotation angle $\Delta \varphi$ between the F and $\mathrm{F}''$ frames, which according to Eq.~\eqref{A31} for $u=\Delta v$ small, is given by
\begin{equation}
\label{A33}
\Delta \varphi \simeq \frac{\Delta v}{v}\Big(\sqrt{1-v^2/c^2}-1\Big)\,.
\end{equation}
As a result, the $\mathrm{O}''$ origin of the $\mathrm{F}''$ frame rotates  with respect to the F frame, with an angular precession velocity $\Omega_{acc}$ along $\textbf{z}$, which for $v/c \ll 1$ reduces, due to Eqs.~(\ref{A32a}) and (\ref{A33}), to
\begin{eqnarray}
\label{A34}
\Omega_{acc} & = &\frac{\Delta \varphi}{\Delta t}\simeq\frac{1}{v} \big(\sqrt{1-v^2/c^2}-1 \big)\frac{\Delta v}{\Delta t}  \nonumber\\
& = & -\frac{1}{v}\Bigg(\frac{1}{\sqrt{1-v^2/c^2}}-1\Bigg) a \simeq -\frac{va}{2c^2}\,.
\end{eqnarray}

The above derivation is done by considering a velocity change $\Delta \textbf{v}=\textbf{u}$ orthogonal to $\textbf{v}$, \textit{i.e.}, an acceleration $\textbf{a}$ orthogonal to $\textbf{v}$. When $\Delta \textbf{v}$ is parallel to $\textbf{v}$, no rotation occurs. This supports the fact that the angular precession velocity for arbitrary $\textbf{v}$ and $\textbf{a}$ has the following form 
\begin{equation}
\label{A35}
\mathbf{\Omega}_{acc}\simeq -\frac{\textbf{v}\times \textbf{a}}{2c^2}\,.
\end{equation}
The rotation reduces to zero when $\textbf{a}=0$ or when $\textbf{a}$ is along  $\textbf{v}$. 

\textbf{(4)} The last step is to use the above results for an electron moving in an external electromagnetic field $(\textbf{H}_{ext}, \textbf{E}_{ext})$. \\
$\bullet$ First, we note that in a magnetic field $\textbf{H}$, a spin-$\textbf{S}$ electron with magnetic moment $\textbf{M}_{S}=g_{e}\frac{e}{2m_{e}c}\textbf{S}$ (see Eq.~\eqref{A3}) rotates with an angular precession velocity 
\begin{equation}
\label{A36}
S \mathbf{\Omega}_{H}= -M_{S}\textbf{H}\,.
\end{equation}
This follows from the fact that the energy $-\textbf{M}\cdot \textbf{H}$ of a magnetic moment $\textbf{M}$ in a magnetic field $\textbf{H}$ gives rise to an interaction term 
\begin{equation}
\label{A37}
\mathcal{W}_{H} =-g_{e}\frac{e}{2m_{0}c} \textbf{S}\cdot \textbf{H}\equiv \omega_{H}S_{z}\,,
\end{equation}
with $\omega_{H}=-Hg_{e}e/2m_{0}c$ for $\textbf{H}$ taken along $\textbf{z}$. The  $\mathcal{W}_{H}$ eigenstates are $\ket{\pm1/2}$ with eigenvalues $\pm \hbar \omega_{H}/2$. Thus, the time evolution of a spin, which is along the $(\theta, \varphi)$ direction when $t=0$ (see Fig.~\ref{fig:anglerot}b), reads as 
\begin{eqnarray}
\label{A38}
 \ket{S_{t}} & = & e^{-i\mathcal{W}_{H}t/\hbar}\left[\cos \frac{\theta}{2}e^{-i\frac{\varphi}{2}}\ket{+} +\sin \frac{\theta}{2}e^{i\frac{\varphi}{2}}\ket{-} \right] \nonumber \\
 & = &  \cos \frac{\theta}{2}e^{-i\frac{\varphi+\omega_{H}t}{2}}\ket{+}+\sin \frac{\theta}{2}e^{i\frac{\varphi+\omega_{H}t}{2}}\ket{-}\,. 
\end{eqnarray}
This shows that $\theta$ does not change with time while $\varphi$ rotates with a velocity $\omega_{H} = -H(M_{S}/S)$ around the $\textbf{z}$ axis,  parallel to $\textbf{H}$,  in agreement with Eq.~\eqref{A36}.\\
 $\bullet$ Next, we note that the spin-$\textbf{S}$ particle located in a frame that moves with a velocity $\textbf{v}$ with respect to the laboratory frame F in which  the electromagnetic field is $(\textbf{H}_{ext}, \textbf{E}_{ext})$, feels a magnetic field $ \textbf{H}'\simeq \textbf{H}_{ext}-\frac{\textbf{v}}{c}\times \textbf{E}_{ext}$. This magnetic field induces an angular precession velocity given, according to Eq.~\eqref{A35}, by
\begin{equation}
\label{A39}
\mathbf{\Omega}_{ext}=- g_{e}\frac{e}{2m_{0}c} \big(\textbf{H}_{ext}+\frac{\textbf{E}_{ext}\times \textbf{v}}{c}\big)\,.
\end{equation}
$\bullet$ The electron also feels in the $\mathrm{F}''$ frame an electrostatic force $e\textbf{E}'\simeq e \textbf{E}_{ext}$ which leads to an acceleration $\textbf{a}$ given by $m_0\textbf{a}=e \textbf{E}_{ext}$. This acceleration brings an additional angular precession velocity given by Eq.~\eqref{A35}. So, we end with
\begin{eqnarray}
\label{A40}
\mathbf{\Omega} & = & \mathbf{\Omega}_{ext}+\mathbf{\Omega}_{acc} \\
& = & -g_{e}\frac{e}{2m_{0}c}\textbf{H}_{ext} - \frac{e}{2m_{0}c}\frac{\textbf{E}_{ext}\times \textbf{v}}{c}(g_{e}-1)\,. \nonumber 
\end{eqnarray}
$\bullet$ According to Eq.~\eqref{A36}, this angular velocity produces an effective magnetic field $\textbf{H}_{eff}=-\mathbf{\Omega} S/M_{S}$, that is, a magnetic energy
\begin{eqnarray}
 \mathcal{W} & = & - \textbf{M}_{S}\cdot \textbf{H}_{eff}=\textbf{S}\cdot \mathbf{\Omega} \label{A41} \\
 & = & -\frac{e}{2m_{0}c} \textbf{S}\cdot\left[g_{e}\textbf{H}_{ext}+(g_{e}-1)\frac{\textbf{E}_{ext}\times \textbf{v}}{c}\right]\,,\nonumber
\end{eqnarray}
in agreement with Eq.~\eqref{A21}. \\
\indent The above derivation, which essentially follows Thomas' idea, has the great advantage to shed light on the physical origin of the spin term appearing in the Dirac equation.
 
 \section {Standard $\textbf{L}\cdot\textbf{S}$ derivation\label{app:B}}
 
\subsection{Spherical harmonics}

 The spherical harmonics $Y_{\ell,\ell_z}(\theta,\varphi)$ for $\ell=1$ read, according to Landau-Lifschitz\cite{LLMQ} phase factor, as
 \begin{eqnarray}
% \label{B.1}
Y_{1,\pm1}(\theta,\varphi)\!\! & = &\!\! \mp i\sqrt{\frac{3}{8\pi}}\sin \theta e^{\pm i\varphi}=i\sqrt{\frac{3}{4\pi}}  \frac{\mp x-iy}{\sqrt{2}r} , \\
Y_{1,0}(\theta,\varphi) \!\!& = & \!\! i\sqrt{\frac{3}{4\pi}}\cos \theta = i\sqrt{\frac{3}{4\pi}}\frac{z}{r} .
\end{eqnarray}
Compared to more common expressions, they contain an additional phase factor $i=e^{i\pi/2}$ that insures $Y_{\ell,\ell_z}(\theta,\varphi)=Y^*_{\ell,-\ell_z}(\theta,\varphi)$, as required from particle-antiparticle symmetry. This particle symmetry is necessary for the consistency of problems that deal with valence holes, like semiconductor excitons.

\subsection{Derivation of the $\ket{j,j_z}$ eigenstates\label{derjjz}}
 By applying $\textbf{J}_{-}=\textbf{L}_{-}+\textbf{S}_{-}$ to
\begin{equation}
\label{}
\ket{j=\frac{3}{2}, j_{z}=\frac{3\eta}{2}}=\ket{\ell=1,\eta}\otimes\ket{\frac{\eta}{2}}\,,
\end{equation}
we find the other two $j=3/2$ states as
\begin{eqnarray}
%\label{B.3}
\ket{j=\frac{3}{2}, j_{z}=\frac{\eta}{2}}= \sqrt{\frac{1}{3  }}\ket{\ell=1,\eta}\otimes\ket{-\frac{\eta}{2}}
\nonumber \\
+ \sqrt{\frac{2}{3  }}\ket{\ell=1,0}\otimes\ket{\frac{\eta}{2}}\,.
\end{eqnarray}

The two $j=1/2$ states made from the same states as $\ket{j=3/2,j_{z}=\eta/2}$ but orthogonal to them, are given by
\begin{eqnarray}
\label{B4}
\ket{j=\frac{1}{2}, j_{z}=\frac{\eta}{2}}=\sqrt{\frac{2}{3  }}\ket{\ell=1,\eta}\otimes\ket{-\frac{\eta}{2}}
\nonumber \\
-\sqrt{\frac{1}{3  }}\ket{\ell=1,0}\otimes\ket{\frac{\eta}{2}}\,,
\end{eqnarray}
within a phase factor irrelevant for the problem at hand.

\subsection{Orbital wave function $\ket{\lambda}$}
The orbital wave functions $\ket{\lambda}$ with $\lambda = (x, y, z)$ defined in  Eq.~(\ref{17}), can be written as
\begin{equation}
\label{B.5}
\braket{\textbf{r}}{\lambda} =\lambda \,f(r)
\end{equation}
with $f(r)=i\sqrt{3/4\pi}R_{n,1}(r)/r$. The orthonormalization of these $\lambda$ states is fulfilled by $f(r)$ such that
\begin{equation}
\label{B.6}
\braket{\lambda'}{\lambda}=\int \mathrm{d}^{3}r \,\lambda'\lambda\,|f(r)|^2=\delta_{\lambda'\lambda}\int \mathrm{d}^{3}r \,x^{2}\,|f(r)|^2\,.
\end{equation} 
Replacing $x^2$ by $y^{2}$ or $z^{2}$ and ultimately by $r^{2}/3$ when the three axes $(x,y,z)$ play the same role as for cubic symmetry, yields
\begin{equation}
\label{B.7}
\braket{\lambda'}{\lambda}=\delta_{\lambda'\lambda}\int r^{2}\mathrm{d}r |R_{n,1}(r)|^2=\delta_{\lambda'\lambda}\,,
\end{equation}
and similarly for the $\ket{n,1,\ell_z}$ states with $\ell_z=(0,\pm1)$.

\subsection{$H_{so}$ eigenstates}
The expressions of the three-fold $p$ orbital states in terms of the $\ket{\lambda}$ states (see Eq.~\eqref{18}), give the two-fold $H_{so}$ eigenstates $\ket{j=1/2, \eta/2}$, given in Eq.~\eqref{23'}, as
\begin{equation}
\label{B.8}
\ket{\frac1 2, \frac\eta 2}= \sqrt{\frac{2}{3}}\,\,\frac{-i\eta\ket{x}+\ket{y}}{\sqrt{2}}\otimes 
\ket{-\frac \eta 2}
-i\sqrt{\frac{1}{3}}\ket{z}\otimes \ket{\frac\eta 2}\,,
\end{equation}
from which we get
\begin{eqnarray}
\lefteqn{ \bra{\frac 1 2, \frac\eta 2}\lambda_{so}(r)\textbf{L}\cdot \textbf{S}\ket{\frac 1 2, \frac\eta 2} } \nonumber \\ 
&&= -\hbar^{2}\int \mathrm{d}^{3}r\,\,\lambda_{so}(r) \,|\braket{\textbf{r}}{1/2,\eta /2}|^{2}\,.
\end{eqnarray}
As the even part of $|\braket{\textbf{r}}{1/2,\eta /2}|^{2}$ is equal to $|f(r)|^{2}(x^{2}+y^{2}+z^{2})/3$, the above quantity reduces to 
\begin{eqnarray}
 \lefteqn{-\frac{\hbar^{2}}{3}\int \mathrm{d}^{3}r\,\lambda_{so}(r)\,r^{2}|f(r)|^{2} } \\
&&= - \hbar^{2}\int_{0}^{\infty}r^{2}{d}r\,\lambda_{so}(r)|R_{n,1}(r)|^{2}  \equiv -2\Lambda_{so}\,.\nonumber 
\end{eqnarray}

In the same way, the four-fold $H_{so}$ eigenstates associated with $j=3/2$ read as 
\begin{eqnarray}
\ket{\frac{3}{2},\frac{3\eta}{2}}\!\! & = & \!\!\frac{-i\eta\ket{x}+\ket{y}}{\sqrt{2}}\otimes \ket{\frac \eta 2}\,, \\
 \ket{\frac{3}{2},\frac{\eta}{2}}\!\! & = &\!\!
\frac{1}{\sqrt{3}} \left[\frac{-i\eta\ket{x}+\ket{y}}{\sqrt{2}}\otimes \ket{-\frac{\eta}{2}}+\sqrt{2}i\ket{z}\otimes\ket{\frac{\eta}{2}}\right]\,. \nonumber \\
\end{eqnarray}
Their eigenvalue is given by $\bra{j=3/2, j_z}H_{so}\ket{j=3/2,j_z}=\Lambda_{so}$.
Therefore, the  spin-orbit interaction $H_{so}$ brings a $3\Lambda_{so}$ splitting between the $(3\times2)$ orbital states with $\ell=1$: four states have an energy shift $\Lambda_{so}$ and two states have an energy shift $-2\Lambda_{so}$.

%%%%%%%%%%%%%%%%%%%%%
% Bibliography
%%%%%%%%%%%%%%%%%%%%

\bibliographystyle{apsrev4-1}
\bibliography{SO_Monique_biblio}

\end{document}